\documentclass[12pt,preprint]{aastex}

\usepackage{graphicx}
\usepackage{tipa}
\usepackage{pdflscape}

\slugcomment{The Astronomical Journal, in press}

\shorttitle{Active Asteroids}
\shortauthors{Jewitt}

\begin{document}

\title{The Active Asteroids}

\author{David Jewitt$^{1,2}$  }
\affil{$^1$ Department of Earth and Space Sciences and \\
$^2$ Department of Physics and Astronomy, UCLA \\
}
\email{jewitt@ucla.edu}

\begin{abstract} 
Some asteroids eject dust, unexpectedly producing transient, comet-like comae and tails.  First ascribed to the sublimation of near-surface water ice,  mass losing asteroids (also called ``main-belt comets'') can in fact be driven by a surprising diversity  of mechanisms.  In this paper, we consider eleven dynamical asteroids losing mass, in nine of which the ejected material is spatially resolved.  We address  mechanisms for producing mass loss including rotational instability, impact ejection, electrostatic repulsion, radiation pressure sweeping, dehydration stresses and thermal fracture, in addition to the sublimation of ice.  In two objects (133P and 238P) the repetitive nature of the observed activity leaves ice sublimation as the only reasonable explanation while, in a third ((596) Scheila), a recent impact is the cause. Another impact may account for activity in P/2010 A2 but this tiny object can also be explained as having shed mass after reaching rotational instability.  Mass loss from (3200) Phaethon  is probably due to cracking or dehydration at extreme ($\sim$1000 K) perihelion temperatures, perhaps aided by radiation pressure sweeping.  For the other bodies, the mass loss mechanisms remain unidentified, pending the acquisition of more and better data.   While the active asteroid sample size remains small, the evidence for an astonishing diversity of mass loss processes in these bodies is clear.
 
\end{abstract}


\section{Introduction}

The classification of small bodies in the inner solar system as either asteroids or comets has historically been attempted by different scientists using different techniques and employing different criteria.  Observational astronomers classify small bodies having transient, unbound atmospheres (usually made visible by the scattering of sunlight from entrained micron-sized dust particles) as ``comets''.  Bodies having instead a constant geometric cross-section are called ``asteroids''.  To planetary scientists, comets and asteroids are distinguished by their ice content or perhaps by their formation location.  Comets are icy (because they formed beyond the ``snow-line'') while asteroids are not (supposedly because they formed at higher mean temperatures inside it).    Lastly, to dynamicists, comets and asteroids are broadly distinguished by a dynamical parameter, most usually the Tisserand parameter measured with respect to Jupiter (Kresak 1982, Kosai 1992).   It is defined by

\begin{equation}
T_J = \frac{a_J}{a} + 2\left[(1-e^2)\frac{a}{a_J}\right]^{1/2}\cos(i)
\label{tisserand}
\end{equation}

\noindent where $a$, $e$ and $i$ are the semimajor axis, eccentricity and inclination of the orbit while $a_J$ = 5.2 AU is the semimajor axis of the orbit of Jupiter.  This parameter, which is conserved in the circular, restricted 3-body problem, provides a measure of the close-approach speed to Jupiter. Jupiter itself has $T_J$ = 3. Main belt asteroids have $a \le a_J$ and  $T_J >$ 3 while dynamical comets have $T_J < 3$.

The three systems of classification (observational, compositional and dynamical) are independent but imperfect.  For example, whether a coma or tail is detected on a given object depends strongly on the parameters of the observing system used.  A puny telescope may not reveal a coma that is easily rendered visible by  a more powerful one.  There is neither an agreed quantitative ice fraction with which to divide comets from asteroids nor, more importantly, any reliable way to measure the ice fraction in a small body.   The formation locations and dynamical histories of small bodies are rendered uncertain by hard-to-model non-gravitational forces from both electromagnetic radiation (the Yarkovsky effect) and mass loss itself (the rocket effect of Whipple 1950) and also by the chaotic nature of solar system dynamics.  The Tisserand criterion is an imperfect classifier because it is based on an idealized representation of the Solar system (e.g.~Jupiter's orbit is not a circle, the gravity of other planets is not entirely negligible, and so on).  Therefore, the utility of Equation (\ref{tisserand}) as a dynamical discriminant is limited for objects with $T_J$ very close to 3.  As an example, the quasi-Hilda comets at $a \sim$ 4.0 AU have $T_J \sim$ 2.9 - 3.04 but are clearly interacting with Jupiter through the  3:2 mean-motion resonance (Toth 2006).  Some recognized Jupiter family comets (e.g. 2P/Encke with $T_J$ = 3.03) also fall in this category (Fernandez et al.~2002, Levison et al.~2006).

\begin{figure}
\epsscale{1.0}
\begin{center}
\plotone{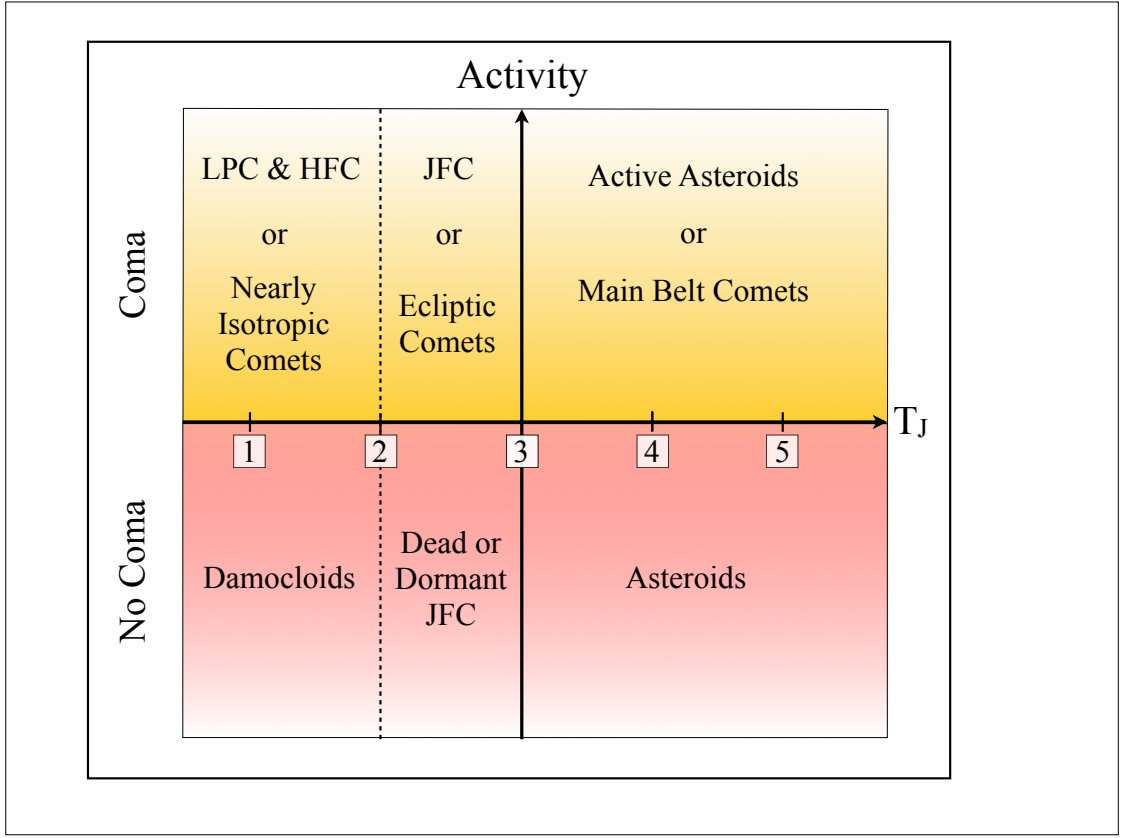}
\caption{Empirical classification of small Solar system bodies by (vertical axis) morphology and (horizontal axis) Tisserand dynamical parameter, $T_J$.  In the figure, LPC = long-period comet, HFC = Halley family comet, JFC = Jupiter family comet.  The dynamical classification using Equation (\ref{tisserand}) assumes $a \le a_J$.  \label{classification} } 
\end{center} 
\end{figure}

Given these and other imperfections it is remarkable that, for a majority of objects, the observational, compositional and dynamical definitions of asteroids and comets lie in close agreement.   For the most part, objects classified as asteroids (comets) based on their orbits, have the physical properties expected of asteroids (comets) as far as can be observed.  Exceptions have given rise to a somewhat confusing and evolving system of nomenclature, used to describe small solar system bodies by a combination of their orbital properties and physical appearances.  To clarify this we show, in Figure (\ref{classification}), a two-parameter classification based on morphology, on the one hand, and the Tisserand parameter on the other.    Traditional comets lose mass and have  $T_J <$ 3.  Those with 2 $\le T_J <$ 3 are called Jupiter family comets and are thought to originate in the Kuiper belt.  Comets with $T_J <$ 2 are long-period or Halley family comets, with a source in the Oort cloud.  Inactive counterparts to the Jupiter family comets are called, variously, extinct, dead or dormant comets (Hartmann et al.~1987).  They are presumably former comets in which past heating by the Sun has removed all near surface ice, although buried ice might remain and these objects could, in principle, reactivate.  Inactive counterparts to the long-period and Halley family comets are called Damocloids (Jewitt 2005).  Again, these are likely objects in which near-surface ice has been lost.  

In this paper, we focus attention on the sub-set of  the small-bodies which are dynamically asteroid-like  ($a < a_J, T_J >$ 3) but which lose mass, like comets.   These were called ``main-belt comets'' by Hsieh and Jewitt (2006) but here we use the term  ``active asteroids'', since some of the examples to be considered, while dynamically asteroid-like and showing comet-like properties, are not in the main-belt.   Numerical integrations show that these are not recently captured comets from the Kuiper belt (Fernandez et al.~2002, Levison et al.~2006).  Scientific interest in these objects lies in the possibility that primordial water ice could have survived in asteroids despite early heating from embedded radioactive nuclei (Grimm and McSween 1989) and heating by the sun. Even greater interest is added by the possibility that the outer asteroid belt may have supplied part of the volatile inventory of the Earth (Morbidelli et al.~2000).  Additionally, active asteroids are a source of dust for the Zodiacal cloud, while unseen counterpart bodies may supply dust to  the debris disks of other stars (e.g. Shannon and Wu 2011).  

After briefly summarizing the current observational evidence concerning these active asteroids, we discuss the surprisingly varied mechanisms through which a body is capable of losing mass.  A recent and complementary discussion focused on observational properties has been offered by Bertini (2011).

\section{Observations of Active Asteroids}

\begin{figure}
\epsscale{0.85}
\begin{center}
\plotone{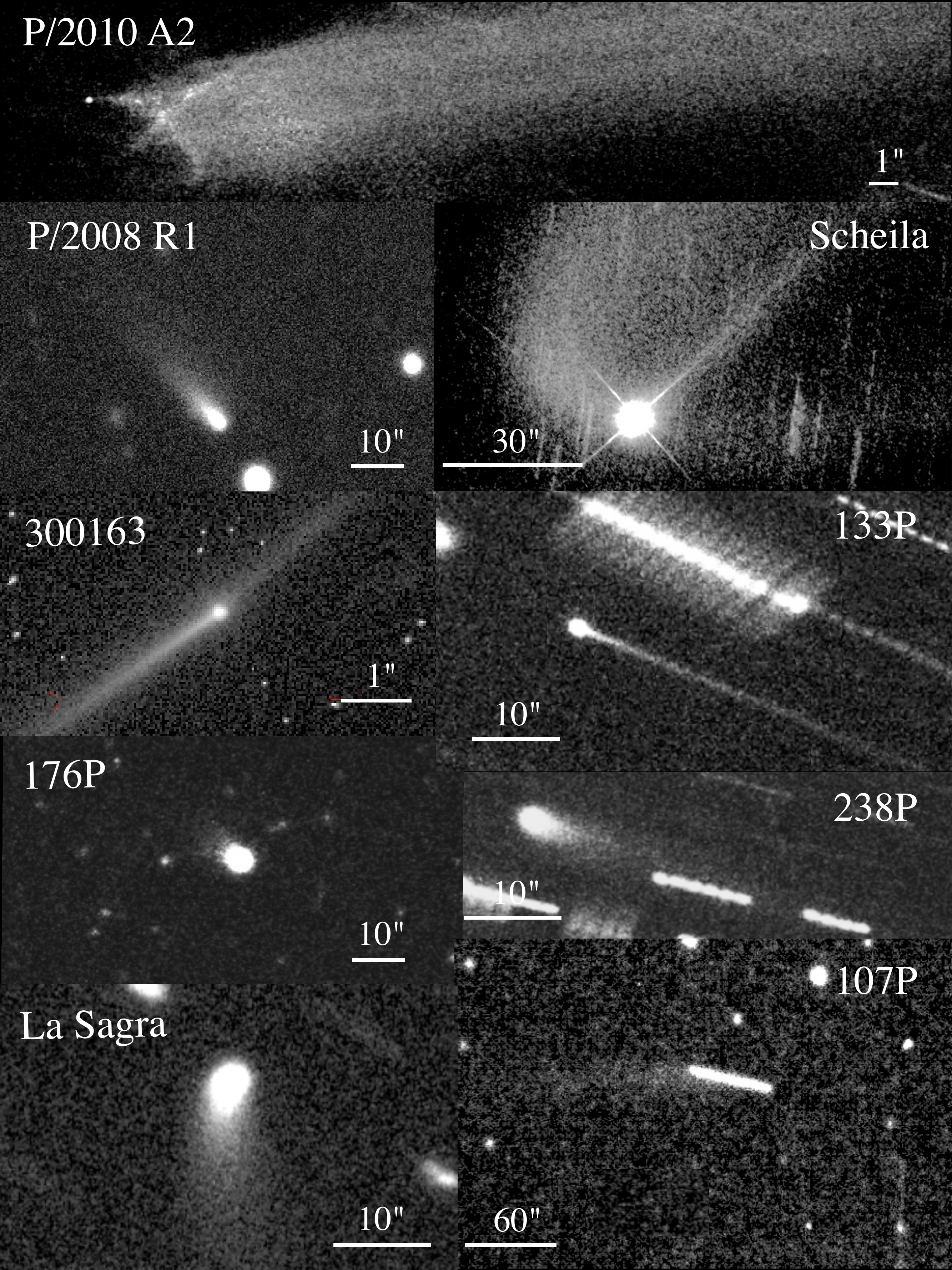}
\caption{Images of nine spatially resolved active asteroids, with scale bars, in the order of decreasing Tisserand parameter, showing their comet-like appearances.  Sources of the images are P/2010 A2 (Jewitt et al.~2010), P/2008 R1 (Jewitt et al.~2009), (596) Scheila (Jewitt et al.~2011), 300163 (Jewitt et al.~2012, in prep.), 133P (Hsieh et al.~2004), 176P (Hsieh et al. 2006), 238P (Hsieh et al. 2009), La Sagra (H. Hsieh, personal communication) and 107P (Fernandez et al.~1997). \label{image_compo} } 
\end{center} 
\end{figure}

\begin{figure}
\epsscale{1.0}
\begin{center}
\plotone{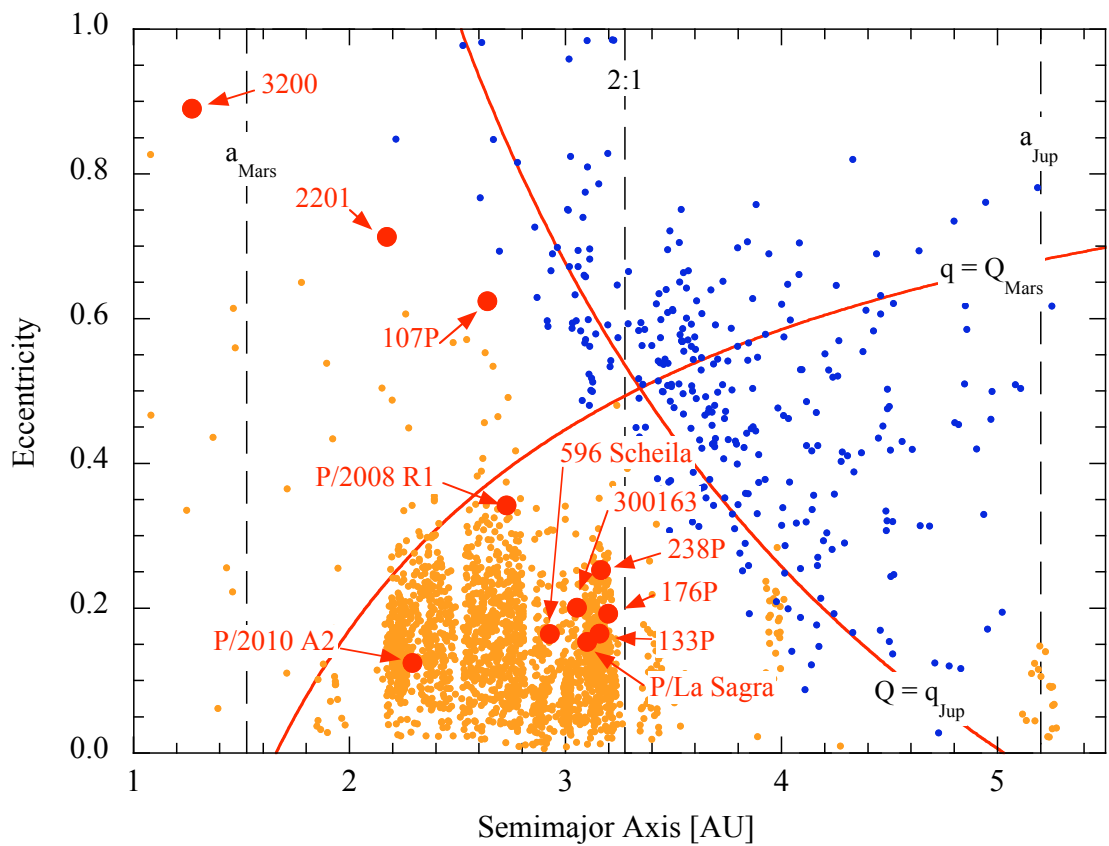}
\caption{Distribution of the active objects (red circles) in the semimajor axis vs.~orbital eccentricity plane.  The corresponding distributions of asteroids (orange circles) and comets (blue circles) are shown for comparison.  Objects above the diagonal arcs cross either the aphelion distance of Mars or the perihelion distance of Jupiter, as marked.  The semimajor axes of the orbits of Mars and Jupiter are shown for reference, as is the location of the 2:1 mean-motion resonance with Jupiter.   \label{mbc_ae_plot} } 
\end{center} 
\end{figure}

To-date, eleven active (or mass-shedding) asteroids  have been reported.  The nine spatially resolved examples are shown for comparison in Figure (\ref{image_compo}), while their positions in the semimajor axis vs.~eccentricity plane are plotted in Figure (\ref{mbc_ae_plot}). The comet-like morphologies and their orbital separation from the domain of the comets are obvious in these two figures.  All eleven objects are located inside the 2:1 mean-motion resonance with Jupiter at 3.3 AU and all but three lie below the Mars-crossing line (marked $q = Q_{Mars}$ in the Figure). The orbital properties of these objects are summarized in Table (\ref{orbital}), the physical properties in Table (\ref{physical}) and we review them briefly here, in order of decreasing $T_J$. 

\textbf{(3200) Phaethon: $T_J$ = 4.508} is dynamically associated with the Geminid meteor stream (e.g. Williams and Wu 1993) and with several other small asteroids including 2005 UD (Ohtsuka et al.~2006, Jewitt and Hsieh 2006, Konoshita et al.~2007) and 1999 YC (Ohtsuka et al.~2008, Kasuga and Jewitt 2008).   All these objects may be related to a precursor body that disintegrated $\sim$10$^3$ yrs ago (Ohtsuka et al.~2006).  Evidence for modern-day mass loss comes from the observed brightening of Phaethon by a factor of two within a few hours of perihelion ($R$ = 0.14 AU)  in 2009 (Jewitt and Li 2010).  This rapid brightening cannot be due to scattering from the $\sim$5 km diameter nucleus alone and, instead, indicates transient mass loss with the ejection of $\sim$10$^8 a_1$ kg of  particles each $a_1$ mm in radius.  With $a_1$ = 1 mm, this is only 10$^{-4}$ of the Geminid stream mass, but raises the possibility that the decay of Phaethon is a continuing process.  The phenomenon has not been observed to repeat, but comparable observations when near perihelion are difficult to secure because of the small angular separation from the Sun.

\textbf{P/2010 A2: T$_J$ = 3.582}
The object showed a distinctive morphology in which a leading, point-like nucleus about 120 m in diameter (Table \ref{physical}), is followed by an extended tail (or trail) of dust in which are embedded ribbon-like structures (Jewitt et al.~2010).  The position angle of the tail and its variation with time are consistent with the action of radiation pressure on mm to cm sized dust particles, following impulsive ejection at very low speeds ($\sim$0.2 m s$^{-1}$) in 2009 February - March (Jewitt et al.~2010, Snodgrass et al.~2010). (Note that an opposite conclusion was reached by Moreno et al.~2010 but based on limited data).   P/2010 A2 went unnoticed for its first $\sim$9 months largely because of its angular proximity to the Sun (Jewitt et al.~2011a).  During this time, a large quantity of fast-moving particles are presumed to have left the vicinity of the main nucleus.  The mass of particles remaining in the tail at discovery is estimated to be in the range (6 to 60)$\times$10$^7$ kg (Jewitt et al.~2010, Moreno et al.~2010, Snodgrass et al.~2010).

\textbf{(2201) Oljato: $T_J$ = 3.299}
Magnetometers on the Pioneer Venus spacecraft revealed multiple, symmetric disturbances in the solar wind magnetic field, clumped non-randomly in time (Russell et al.~1984).  About 25\% of these events are associated with planet-crossing asteroid (2201) Oljato, whose orbit lies interior to Venus' when near perihelion.  Russell et al. suggested that the magnetic disturbances result from decelleration of the solar wind, perhaps caused by mass loading from ionized gases released by an unknown process from debris distributed along Oljato's orbit.  A mass loading rate of only $\sim$5 kg s$^{-1}$ is reportedly needed.  
However, a spectroscopic search for gas produced by Oljato itself proved negative (Chamberlin et al. 1996), with upper limits to the CN production rate near 10$^{23}$ s$^{-1}$.  With a standard H$_2$O/CN mixing ratio of 360, the corresponding limit to the mass production rate in water is $\le$1.5 kg s$^{-1}$.  Whatever the cause of the repetitive magnetic disturbances, they are not products of an inert asteroid and imply mass loss from Oljato. A dynamical simulation indicates that (2201) Oljato has negligible chance of being a captured Jupiter family comet (Bottke et al.~2002).

\textbf{P/2008 R1: T$_J$ = 3.216}
This object was observed over a $\sim$45 day interval in 2008, when near $R$ = 2 AU, as having the appearance of an active comet with a typically flared tail (Jewitt et al.~2009).  The object intrinsically faded by a factor of about 2 over the above interval of observations.  An upper limit to the nucleus radius of 0.7 km was set (red geometric albedo 0.05 assumed, see Table \ref{physical}), later reduced to $r_n \le$ 0.2 km by subsequent observations (H. Hsieh, private communication, 2011).  Spectral observations limited the production of the CN radical to Q$_{CN} \le$ 1.4$\times$10$^{23}$ s$^{-1}$, again corresponding to a water production rate $\le$1.5 kg s$^{-1}$ assuming H$_2$O/CN = 360.  P/2008 R1 is located near the 8:3 mean-motion resonance with Jupiter and is also affected by the $\nu_6$ secular resonance.   The dynamical lifetime in this orbit is short (20 to 30 Myr) compared to the age of the solar system, suggesting that P/2008 R1 was scattered into its present location from elsewhere in the asteroid belt.

\textbf{(596) Scheila: T$_J$ = 3.208}
(596) Scheila, a 113 km diameter object with red geometric albedo $\sim$0.04 (Table \ref{physical}), developed a coma in late 2010.  Over the course of a month, this coma expanded with a characteristic speed $\sim$60 m s$^{-1}$ and faded in response to the action of solar radiation pressure (Bodewits et al.~2011, Jewitt et al.~2011, Moreno et al.~2011), apparently without any continued replenishment of particles from the nucleus.  The gas production from the nucleus was reportedly limited to Q$_{OH} \le$10$^{25}$ s$^{-1}$ (Moreno et al.~2011) to Q$_{OH} \le$10$^{26}$ s$^{-1}$ (Howell and Lovell 2011), corresponding to water production rates $\le$0.3 kg s$^{-1}$ to 3 kg s$^{-1}$ (the meaning of these limits is unclear given the non-steady nature of the mass loss event from Scheila).  The mass of dust in micron-sized grains was  4$\times$10$^7$ kg (Jewitt et al.~2011) while more model-dependent attempts to account for mass in larger particles gave 6$\times$10$^8$ kg (Bodewits et al.~2011) to 2$\times$10$^{10}$ kg (Moreno et al.~2011).  No ice was observed in the coma (Yang and Hsieh 2011).  

\textbf{300163 (2006 VW139): T$_J$ = 3.203}
Discovered in 2006 and first observed to be active on UT 2011 Aug. 30 (Hsieh et al.~2011d), little is yet known about this object.  With reported absolute magnitude $H$ = 16.6 and an assumed albedo $p_V$ = 0.04, appropriate to its outer belt location, the estimated diameter is $\sim$3 km.  Two thin tails, one near the projected orbit and another roughly antisolar, show that dust leaves the nucleus very slowly and point to mass loss over a protracted period.  There is no reported evidence for gas or for repetitive mass loss, but new observations of this object are still be acquired at the time of writing.

\textbf{133P/Elst-Pizarro: T$_J$ = 3.184}
The appearance is typically that of a point-like nucleus with a thin tail (or ``trail'') of 10 $\mu$m sized dust following in the projected orbit.  Order of magnitude mass loss rates in gas, inferred from spectroscopy (Licandro et al.~2011) and  in dust, inferred from surface photometry of the tail (Hsieh et al.~2004) reach $\sim$0.02 kg s$^{-1}$.  The thin tail indicates that particles are ejected from the nucleus at speeds $\sim$1.5 m s$^{-1}$, comparable to the escape speed from the 3.8$\pm$0.6 km diameter nucleus.

\textbf{176P: T$_J$ = 3.167}
A cometary appearance was detected for 176P/LINEAR over a single, month-long interval in 2005 (Hsieh et al.~2011a).  During this time, the object showed a fan-shaped tail and was about 30\% brighter than the bare nucleus, leading to an implied dust mass $\sim$10$^5$ kg.  The properties of the dust can be approximately matched by models in which the characteristic particle size is 10 $\mu$m, the ejection speed $\sim$5 m s$^{-1}$ and the dust production rate $\sim$ 0.07 kg s$^{-1}$, all similar to values inferred in 133P.  The 4.0$\pm$0.4 km diameter nucleus rotates with a period near 22.2 hr (Table \ref{physical}).

\textbf{238P/Read (= P/2005 U1): T$_J$ = 3.152}
Like P/2008 R1, the nucleus of 238P/Read is tiny, with a diameter $\sim$0.8 km (Hsieh et al.~2011b).  It was observed to be in an active state in both 2005 and 2010, but not in between, with a coma dust mass of order 10$^5$ kg and a production rate estimated (from published photometry) near $\sim$0.1 kg s$^{-1}$.  Also like P/2008 R1, 238P/Read is dynamically short-lived, with a survival time of order 20 Myr (Haghighipour 2009).

\textbf{P/2010 R2 (La Sagra): T$_J$ = 3.098}
The object was observed to be active from 2010 September to  2011 January, at $R$ = 2.6 to 2.7 AU.   Moreno et al.~(2011b) inferred dust production at the peak rate of $\sim$4 kg s$^{-1}$, with centimeter-sized particles ejected at about 0.1 to 0.2 m s$^{-1}$.  A limit to the outgassing rate $Q_{CN} \le$3$\times$10$^{23}$ s$^{-1}$ (corresponding to 3 kg s$^{-1}$ in water) was placed spectroscopically (Hsieh et al.~2011c). Hsieh et al.~(2011c) show (neglecting possible non-gravitational forces due to outgassing) that the orbit of P/2010 R2 is stable on timescales $\sim$100 Myr and argue that this object was likely formed in-situ.  

\textbf{107P/(1949 W1) Wilson-Harrington: T$_J$ = 3.083}
Fernandez et al.~(1997) analyzed two photographic plates taken on 1949 November 19, when at $R$ = 1.148 AU.  The object is trailed in both but, in the blue plate, shows a prominent diffuse tail about 2$\arcmin$ in length.  The red plate shows only a hint of this tail.  The color (B-R = -1) is inconsistent with scattering from dust, but suggests instead resonance fluorescent scattering from an ion tail.  The position angle of the tail, being about 15$\degr$ from radial to the sun, is also more consistent with the expected direction of a plasma tail than with a dust tail.  Curiously, 107P was re-observed on November  22 and 25 but then showed no trace of a tail (Cunningham 1950), and no comet-like activity has been reported since (Lowry and Weissman 2003, Ishiguro et al.~2011).  Based on a statistical dynamical model, Bottke et al.~(2002) concluded that there is a 4\% chance that 107P  is a captured Jupiter family comet.

\section{Mass Loss Mechanisms}
\label{mechanisms}

We consider a variety of processes  capable of launching dust from a small body.  In each case, the number of unknown but relevant physical parameters prevents any exact treatment, but it remains instructive to consider the range of action of the mechanisms in the context of the asteroids.

\subsection{Sublimation}
Since Whipple (1950), sublimation has been explored in great detail as the driver of mass loss from the classical comets. It need not be re-described in detail here. Although simple in concept, detailed studies of comets show that sublimation is a remarkably complex process when factors such as the porosity of the surface, nucleus rotation, the conduction of heat into the interior and the development of a refractory mantle are considered (Guilbert-Lepoutre et al.~2011).  One simplification possible for the present objects is the assumption that asteroids contain no amorphous ice, since temperatures in the asteroid belt are too high for it to escape crystallization.  Accordingly, we address only the highly idealized case of sublimation from an exposed crystalline ice surface in thermal equilibrium with sunlight.

The sublimation mass flux per unit area, $dm/dt$ (kg m$^{-2}$ s$^{-1}$) from a patch of surface whose normal is inclined to the Sun-direction by angle $\theta$ is determined by solution of the energy balance equation

\begin{equation}
\frac{F_{\odot} (1-A)}{R_{AU}^2} \cos(\theta) = \epsilon \sigma T^4 + L(T) \frac{dm(T)}{dt} + f_c.
\label{subl}
\end{equation}

\noindent Here $F_{\odot}$ = 1360 W m$^{-2}$ is the Solar constant, $R_{AU}$ is the heliocentric distance in AU, $A$ is the Bond Albedo, $\epsilon \sim$ 0.9 is the emissivity of the surface, $\sigma$ = 5.67$\times$10$^{-8}$ W m$^{-2}$ K$^{-4}$ is the Stefan-Boltzmann constant, $T$ is the equilibrium temperature and $L(T)$ is the latent heat of sublimation.  The temperature dependence of $dm/dt$ may be estimated from thermodynamics and the Clausius-Clapeyron equation, or from direct measurement (Washburn 1926).  The conduction term, $f_c$, requires knowledge of the thermophysical parameters and the spin vector of the nucleus and is small. We neglect it here.

The angle term, $\cos(\theta)$, depends on the nucleus rotation vector, changes with the position on the nucleus and varies with time as the nucleus rotates.  Therefore, the  total mass loss rate must be obtained by integrating Equation (\ref{subl}) over the surface and with respect to time.  To avoid sinking in complexity, we here focus on two limiting cases for the surface temperature and resulting sublimation rate.  The maximum temperature, $T_{max}$, is reached at the subsolar point on a non-rotating body, where 
$\cos (\theta)$ = 1.  The minimum effective temperature, $T_{min}$, occurs when the surface is isothermal, corresponding to the largest possible radiating area.  On a sphere, the average value of the angle term is  
$\overline{\cos (\theta)}$ = 1/4.  Accordingly, we estimate the minimum and maximum temperatures and specific sublimation rates from Equation (\ref{subl}) for $\overline{\cos (\theta)}$ = 1 and 1/4.

\begin{figure}
\epsscale{0.9}
\begin{center}
\plotone{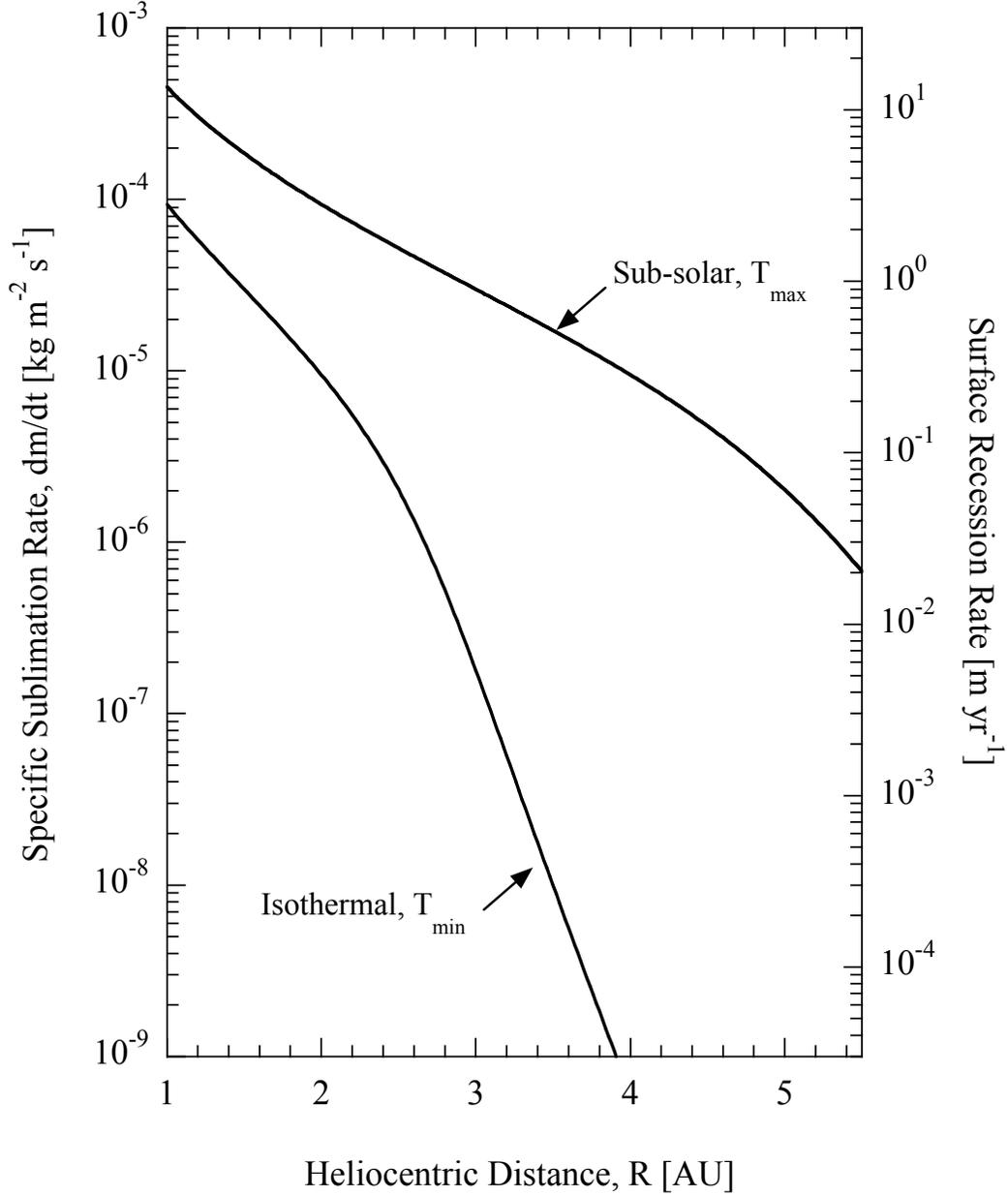}
\caption{On the left, the equilibrium water ice sublimation rate (from Equation \ref{subl}) and, on the right, the surface recession rate (from Equation \ref{recession}), both as functions of the heliocentric distance.  The curves are for albedo 0.05 and the maximum and minimum equilibrium temperatures, corresponding to the subsolar point on a non-rotating body and to an isothermal surface, respectively.   \label{sublimation} } 
\end{center} 
\end{figure}

We considered both low ($A$ = 0.05) and high ($A$ = 0.50) ice albedos and calculated sublimation mass fluxes from Equation (\ref{subl}).    Clean (high albedo) ice sublimates too slowly to drive activity at asteroid belt distances and so we present only the dirty (low albedo) ice solutions for the sublimation rate as a function of distance, $R$,  in Figure (\ref{sublimation}).  The curves are convergent towards small $R$ because the sublimation term in Equation (\ref{subl}) is then dominant, but separate dramatically at larger $R$ as the sublimation term drops exponentially, leaving all the energy for radiation.   

Across the 2.0 $\le R \le$ 3.3 AU strip corresponding to the asteroid belt, sublimation rates from $\sim$10$^{-4}$ kg m$^{-2}$ s$^{-1}$ (sublimation at $T_{max}$) to $<$10$^{-8}$ kg m$^{-2}$ s$^{-1}$ (sublimation at $T_{min}$)  are found.  Mass loss rates near 1 kg s$^{-1}$ (Table \ref{physical}) require sublimating areas from 10$^4$ m$^2$ ($T_{max}$) to 10$^8$ m$^2$ ($T_{min}$).   Since the latter exceeds the entire surface area of (for example) 133P, we can exclude the isothermal ($T_{min}$) model as the activity driver.   The main conclusion to be drawn from Figure (\ref{sublimation}) is that exposed water ice can sublimate across the full range of heliocentric distances occupied by the main belt provided it is dirty.  

Two other quantities are directly calculable once $dm/dt$ is known.  First, the rate of recession of the sublimating surface is given by 

\begin{equation}
\frac{d \ell}{dt} = \frac{1}{\rho} \frac{dm}{dt}
\label{recession}
\end{equation}

\noindent and indicated on the right-hand axis of Figure (\ref{sublimation}).  With ice density $\rho$ = 1000 kg m$^{-3}$, we find 350 $\mu$m yr$^{-1}$ $\le d\ell/dt \le$ 3 m yr$^{-1}$ over the 2 to 4 AU distance range swept by the majority of active asteroids (see right hand axis of the Figure).

\begin{figure}
\epsscale{0.9}
\begin{center}
\plotone{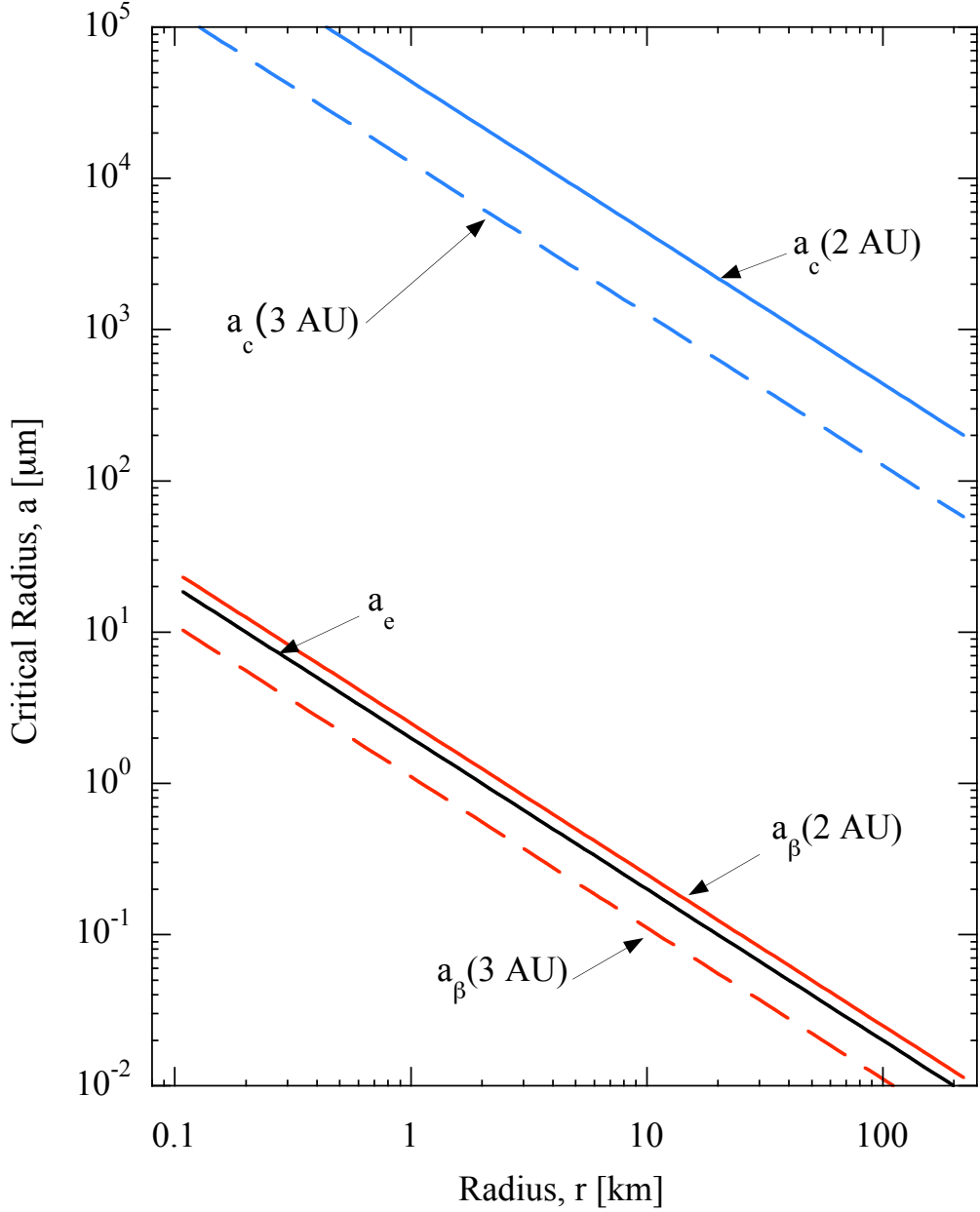}
\caption{Critical dust grain radius for ejection by gas drag ($a_c$, computed from Equation \ref{drag}), for electrostatic ejection ($a_e$, Equation \ref{electro}) and for  loss by radiation pressure sweeping ($a_{\beta}$, Equation \ref{amu})  as functions of nucleus radius.  Solid and dashed curves for  $a_c$ and $a_{\beta}$ refer to $R$ = 2 AU and 3 AU, as marked.    \label{critical} } 
\end{center} 
\end{figure}

Second, the size of the largest ejectable particle is estimated by balancing the gas drag force with the gravity, assuming spherical particles and nucleus and neglecting all other forces.  We take the densities of the ejected grains and of the bulk nucleus, in the absence of contrary evidence,  to be equal. The resulting maximum particle radius, computed from

\begin{equation}
a_c = \frac{9}{16 \pi}  \frac{C_D V_{th}}{G \rho^2 r}\frac{dm}{dt}
\label{drag}
\end{equation}

\noindent is plotted in Figure (\ref{critical}) as a function of heliocentric distance for low albedo, subsolar sublimation.  In Equation (\ref{drag}), $C_D$ is the dimensionless drag coefficient, here taken to be unity, and $V_{th}$ is the gas speed at the nucleus surface, taken from Biver et al.~(2002).  The equation offers, at best, a crude estimate of the critical radius; the highly anisotropic outflow of gas from a real nucleus can strongly affect dust acceleration and dynamics relative to the simple isotropic case used here (e.g. Crifo et al.~2005).  Still, it is obvious from Figure (\ref{critical}) that sublimation gas drag can easily remove small solid particles across the full range of distances swept by the main-belt asteroids.  Grains larger than $a_c$ remain bound to the asteroid and contribute to the development of a refractory mantle, eventually leading to a stifling of the sublimation.

While exposed ice sublimates quickly, ice protected from direct illumination by a covering or mantle of refractory debris can survive nearly indefinitely.  A meter-thick layer of regolith can suppress sublimation sufficiently to retain ice for Gyr timescales (Schorghofer 2008).  Then, to initiate sublimation, this protective mantle must be removed, which could occur naturally through impact.  

\subsection{Impact Ejection}
Collisions between asteroids occur at characteristic speeds of several km s$^{-1}$ (Bottke et al.~1994) and are therefore highly erosive.  Impact yields, defined as the ratio of the ejecta mass, $m_e$, to the projectile mass, $M$, are $m_e/M \gg$1.  Material ejected by impacts may explain observed activity in some asteroids.  In this section we aim to obtain a relation between the impactor properties and the resulting brightening caused by ejected material.

In an impact, the bulk of the ejecta travel at the lowest speeds.  For equal target and projectile densities, the mass of ejecta, $m_e$, traveling faster than a given speed, $v$, can be roughly expressed by a power law 

\begin{equation}
m_e/M= A (v/U)^{\alpha}
\label{housen}
\end{equation}

\noindent in which $M$ and $U$ are the impactor mass and speed and $A \sim$ 0.01 is a constant (Housen and Holsapple 2011).  The index $\alpha$ depends slightly on the properties of the target but is reasonably well approximated for a range of materials as $\alpha$ = -1.5.  Only ejecta traveling with $v \ge v_e$ can escape from an asteroid to produce an increase in the scattering cross-section, while the rest must fall back to coat the surface around the impact site.  With the escape velocity given by

\begin{equation}
v_e = \left(\frac{8 \pi G \rho}{3} \right)^{1/2} r
\label{escape}
\end{equation}

\noindent it is clear that  Equation (\ref{housen}) defines, for a given target density $\rho$ and  impact speed $U$, a relation between the impact yield and the radius, $r$, of the impacted asteroid.

A relation between $m_e$ and the scattering cross-section of the ejecta and hence the change in brightness caused by impact, can be established.  The size distribution of the ejecta from NASA's Deep Impact mission to comet 9P/Tempel 1 has been modeled as a power law (Kadono et al.~2010). Their Figure 4 gives a differential power law size index $q$ = 3.7 over the 1 $\le a \le$ 100 $\mu$m size range, while other investigators have found slightly steeper distributions (Lisse et al.~2006).  Separately, the size distribution of the ejecta from P/2010 A2 has been modeled as a power law with $q$ = 3.3$\pm$0.2 for $1 \le a \le 10$ mm (Jewitt et al.~2010).  A wider range of power laws (from $q$ = 3 to as steep as $q$ = 6) has been reported in small-scale, hypervelocity impact experiments (Takasawa et al.~2011).  For the sake of the present discussion, we adopt $q$ = 3.5 and find that the relation between the scattering cross-section, $C_e$, and the mass of particles having radii in the range $a_{min} \le a \le a_{max}$ is

\begin{equation}
m_e = \frac{4}{3} \rho \overline{a} C_e
\label{mvsC}
\end{equation}

\noindent where $\overline{a} = (a_{min} a_{max})^{1/2}$.  For example, with $a_{min}$ = 0.1 $\mu$m, $a_{max}$ = 0.1 m, $\overline{a}$ = 10$^{-4}$ m, or 0.1 mm.  

We combine Equations (\ref{housen}), (\ref{escape}) and (\ref{mvsC}) to compute the ratio of the cross-section of the ejecta to the cross-section of the target asteroid


\begin{equation}
\frac{C_e}{\pi r^2}=  \frac{A }{\overline{a} U^{\alpha} } \left(\frac{8 \pi G \rho}{3} \right)^{\alpha/2}
r_p^3 r^{\alpha - 2}
\label{dm}
\end{equation}

\noindent where we have taken the projectile to be a sphere of the same density as the target and with a radius, $r_p$.  Substituting $A$ = 0.01, $\overline{a}$ = 0.1 mm, $U$ = 5 km s$^{-1}$, $\rho$ = 2000 kg m$^{-3}$ and $\alpha$ = -3/2, we obtain

\begin{equation}
\frac{C_e}{\pi r^2} \sim 30 \left(\frac{r_p}{1 m}\right)^{3}\left(\frac{r}{1 km}\right)^{-7/2}
\label{ratio}
\end{equation}

\noindent A 1 m projectile impacting a 1 km target asteroid will produce ejecta with a cross-section 30 times the geometric cross-section of the asteroid.  

\begin{figure}
\epsscale{0.85}
\begin{center}
\plotone{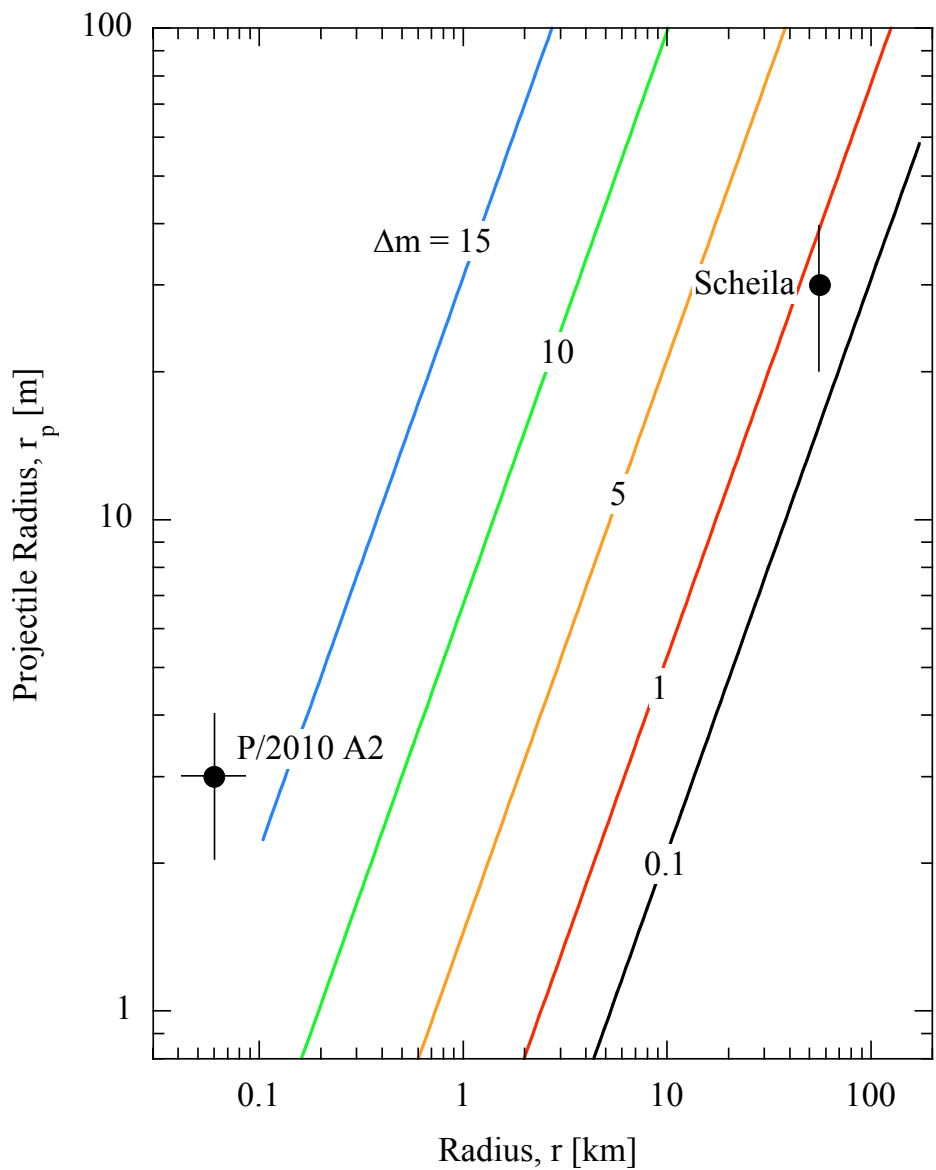}
\caption{Lines of constant brightening, in magnitudes, as a function of target and impactor radius.  Impact velocity U = 5000 ms$^{-1}$ and ejecta with a $q$ = 3.5 differential size power law with particles in the 0.1 $\mu$m to 0.1 m size range were assumed.  The radii of P/2010 A2 and (596) Scheila and their presumed impactors are shown, with error bars (the radius uncertainty for P/2010 A2 assumes a factor-of-two error in the assumed albedo). \label{dmag} } 
\end{center} 
\end{figure}

Lines of constant brightening, computed from

\begin{equation}
\Delta m = 2.5 \log_{10}\left[1 + \frac{C_e}{\pi r^2}\right]
\label{deltamag}
\end{equation}  

\noindent are shown in  Figure (\ref{dmag}), as a function of asteroid and projectile radius.  The curves are plotted only where the ejecta mass computed by Equation (\ref{housen}) is less than the target mass.  The Figure shows that, for example, the impact of a 1 m projectile into a 1 km asteroid would cause a brightening by $\Delta m \sim$ 3.5 magnitudes, while impact of the same projectile into a 10 km asteroid would cause a brightening of only a few $\times$0.01 magnitude, and would almost certainly escape detection.  The published parameters of (596) Scheila and P/2010 A2, interpreted as impact ejecta, are also plotted on Figure (\ref{dmag}).  The location of Scheila overlaps the $\Delta m$ = 1 magnitude brightening line, consistent with observations.  P/2010 A2 was discovered nearly a year after disruption. The magnitude of its initial brightening is thus unknown, but Figure (\ref{dmag}) suggests that it was a dramatic $\Delta m\sim $ 15.   The curves in Figure (\ref{dmag}) are uncertain, both because of the simplicity of the treatment and because of the many uncertainties in the impactor and dust grain parameters.  For example, if $U$ were larger by a factor of two, Equation (\ref{dm}) shows that the ejecta mass and cross-section would be larger by a factor of $\sim$2.8 and the ejecta would be brighter by $\sim$1 magnitude.  Nevertheless, Equation (\ref{dm}) and Figure (\ref{dmag}) provide a useful first estimate of the observable effect of a given impact in the main belt.   The Figure leads naturally to the question of the impact rate in the main-belt, which we discuss in Section (\ref{discussion}).

While some ejecta can be launched from the impact site at speeds comparable to the impact speed (i.e. $v \sim U \sim$ 5 km s$^{-1}$), most material travels much more slowly, following Equation (\ref{housen}).  The slowest escaping material leaves the target at a speed comparable to the gravitational escape speed, thus setting a relation between the size of the target and the duration of the brightening caused by impact.  For example, consider two target asteroids of size 100 m and 100 km, observed from 1 AU distance using a typical charge-coupled device camera with a field of view 5$\arcmin$ (corresponding to $\sim$10$^5$ km half-width).  Ejecta from the  100 km body will travel at $\sim$100 m s$^{-1}$ and cross the field of view in $\sim$10$^6$ s (about 10 days).  Ejecta from the 100 m body will travel at characteristic speed $v_e \sim$ 0.1 m s$^{-1}$ and take so long to cross the field ($\sim$10$^9$ s) that radiation pressure and Keplerian shear will dominate the distribution of the dust.  The persistence of impact-produced coma is thus a measure of the size of the target body.

\subsection{Rotational Instability}
The equatorial, centripetal acceleration on a strengthless rotating object of mass density $\rho$ (kg m$^{-3}$) equals the gravitational acceleration at the critical rotational period, $P_c$, given by

\begin{equation}
P_c = k \left[\frac{3 \pi}{G \rho}\right]^{1/2}.
\label{rotation}
\end{equation}

\noindent Here, $k$ is a dimensionless constant dependent on the shape of the object.  A  sphere has $k$ = 1. A prolate body with axes $a \ge b$ in rotation about its shortest dimension has $k \sim a/b$.  Expressed in hours, Equation (\ref{rotation}) may be written

\begin{equation}
P_c~[hr] \sim 3.3 k  \left[\frac{1000}{\rho}\right]^{1/2}.
\label{rotation2}
\end{equation}

\noindent For example,  a sphere with $\rho$ = 2000 kg m$^{-3}$  has $P_c$ = 2.3 hr while an elongated body with the same density and $a/b$ = 2  has $P_c$ = 4.7 hr, by Equation (\ref{rotation2}).  Although Equation (\ref{rotation2}) shows that  $P_c$ is independent of the object size,  numerous measurements of  asteroids show that only objects smaller than $\sim$0.1-0.3 km rotate with periods $<$2 hr (Pravec et al.~2002).  The probable reason is that sub-kilometer asteroids have significant tensile strength, and so can resist rotational disruption, while larger bodies are structurally weak and rotationally unstable when $P \le P_c$.  A weak (in tensile strength) aggregate or ``rubble pile'' structure could be produced by impact fracturing or, in the case of cometary nuclei, by the gentle settling together of icy planetesimals formed separately.

Given the evidence from the distribution of asteroid rotation rates, it is reasonable to conjecture that rotational instability might be a cause of observable mass-loss in the main-belt.  A number of  torques are capable, in principle, of driving an asteroid into rotational instability.  For example, collisions between asteroids lead to a random walk in angular momentum that can end in an unstable state.  Anisotropic radiation of photons can lead to a net torque (the so-called YORP effect) that can do the same (Marzari et al.~2011, Jacobson and Scheeres 2011).  An object losing mass through sublimation can, depending on the rate and angular dependence of the mass loss, experience a torque orders of magnitude larger than torques due to either collisions or YORP, quickly being driven towards rotational break-up (c.f. Drahus et al.~2011).  While, in principle, rotational instability can affect any asteroid regardless of its size, in practice the spin-up times vary inversely with radius and the objects most likely to experience rotational bursting are the smallest (exactly as observed in the near-Earth population).  Could some active asteroids be rotationally disrupted objects?

\subsection{Electrostatic Forces}
The action of electrostatic forces in moving dust particles across a planetary surface is best established on the Moon.  Images from the Surveyor lunar lander spacecraft  showed an unexpected ``horizon glow'',  caused by forward scattering from dust particles $\sim$10 $\mu$m in size  located $\sim$1 m above the surface (Rennilson and Criswell 1974).  Later,  the Lunar Ejecta and Meteorites (LEAM) dust detection experiment of Apollo 17 recorded the impact of low velocity ($\le$100 m s$^{-1}$) dust particles, and showed particularly intense fluxes of such particles near local sunrise and sunset (Berg et al.~1976).  The measured dust particles fluxes are $\sim$7 orders of magnitude too large to be associated with churning of the surface by micrometeorite bombardment.  

Instead, electrostatic levitation was suggested as the source of these particles.  On the day-side, positive charging of the lunar surface occurs by ejection of electrons owing to the photoelectric effect from UV and X-ray solar photons.  In the absence of a discharging current, the potential can rise to the energies of the most energetic photons (10$^3$ V, or more, c.f. De and Criswell 1977) but, in practice, the measured dayside potential is about +10 V (Colwell et al.~2007). On the unilluminated side, or in shadows, the surface becomes negatively charged as a result of the greater flux of solar wind electrons (the solar wind is electrically neutral, but the low-mass electrons travel much faster than solar wind protons, leading to a net flux of electrons onto the surface).  The typical energy of solar wind electrons is $\sim$10 eV leading to a negative potential of order 10 V.  Near the terminator, the same processes lead to positive and negative charging of  the surface, but on much smaller spatial  scales corresponding to the scale of the local topography, leading to locally high electric field gradients.   Near a shadow edge, the photoelectrons produced in sunlight can travel to and stick in unilluminated regions, causing local electric field gradients estimated at $E \sim$ 10 to 100 V m$^{-1}$ (Colwell et al.~2007).  The positive, sunlit surface attracts a cloud of electrons, effectively neutralizing the gradient on length scales $\ell \gtrsim$1 m.

The electrostatic processes that move dust particles on the Moon presumably operate also on the asteroids.  The charging time on the Moon is $\sim$10$^2$ to 10$^3$ s (de and Criswell 1977); photoelectron charging currents will be 9 times weaker at 3 AU and the charging times will be 9 times longer but the potentials attained, for a given dielectric constant, will remain the same.  The principal difference in the asteroid belt is that, whereas levitated lunar dust is retained by the gravity of the Moon, dust ejection speeds on small bodies can exceed the escape velocity, $v_e$.  Therefore, electrostatic processes are potentially capable of leading to mass loss from asteroids.    

The charge on a spherical grain of radius $a$ is related to the potential on the grain, $V$, by $q$ = $4\pi\epsilon_0 V a$, where $\epsilon_0$ = 8.854$\times$10$^{-12}$ F m$^{-1}$ is the permittivity of free space.  The force on a charged particle exposed to an electric field $E$ (V m$^{-1}$) is just $F_e = q E$.  As our criterion for dust ejection, we demand $v > V_e$, where $v$ is the grain speed achieved by accelerating across the shielding distance $\ell$ and $V_e$ is the gravitational escape speed at the surface.  Assuming that the grain and the asteroid are spherical, of radius $a$ and $r$, respectively, and that both have density $\rho$, this criterion gives the critical grain size for electrostatic ejection as

\begin{equation}
a_e= \left(\frac{18 \epsilon_0 V E \ell}{4 \pi G \rho^2 r^2}\right)^{1/2}.
\label{electro}
\end{equation}

Substituting the lunar values,  $V$ = 10 Volts, $\ell$ = 1 m, $E$ = 10 to 100 V m$^{-1}$, and using $\rho$ = 2000 kg m$^{-3}$ as the canonical asteroid density, Equation (\ref{electro}) gives $a_e$ = 1.5 to 5 $\mu$m for a $r$ = 1 km asteroid (c.f. Figure \ref{critical}).  These sizes are somewhat smaller than the $\sim$10 $\mu$m particle sized particles inferred from observations of some active asteroids (e.g. 133P; Hsieh et al. 2004) but, given the many uncertainties in both the model and in the interpretation of observations, perhaps the differences are acceptable.  In contrast, Equation (\ref{electro}) shows that on the Moon ($r$ = 1600 km), only nanometer-sized grains can be ejected, meaning that the process is irrelevant there. Even for asteroid (596) Scheila ($r$ = 56 km and $a_e$ = 0.04 to 0.1 $\mu$m), any electrostatically ejected particles would be smaller than a wavelength and inefficient optical scatterers.  Figure (\ref{critical}) shows that $a_e \sim 10^{-4} a_c$, for a given asteroid radius.  We conclude that electrostatic ejection of particles large enough to scatter optical photons is a plausible mass-loss mechanism only for smaller asteroids.

There are major uncertainties concerning electrostatic effects on the Moon, especially regarding the tendency for small particles to stick to each other and to surfaces, through the action of Van der Waals and other contact forces.  These uncertainties are magnified in importance on the asteroids  because, unless the forces of cohesion can be overcome, electrostatic levitation and ejection will be impossible (Hartzell and Scheeres 2011). Unfortunately, ignorance of grain cohesion limits our ability to know whether electrostatic ejection is important on the asteroids, although flat, pond-like structures on Eros have been interpreted this way (Hughes et al.~2008).  Sonnett et al.~(2011) report that $\sim$5\% of main-belt asteroids show ultra-faint comae that cannot be detected individually but which are collectively significant.  If these comae are real, electrostatic ejection of sub-micron grains might offer a plausible source mechanism that is approximately independent of distance from the sun.


\subsection{Thermal Fracture}
Thermal fracture can occur when the stress associated with a change in the temperature of a material exceeds the tensile strength of the material.  The thermal stress is
$S \sim \alpha Y \delta T$, where $\alpha$ (K$^{-1}$) is the thermal expansivity, $Y$ (N m$^{-2}$) is the Young's Modulus and $\delta T$ (K) is the responsible temperature change.  The expansivities of common rocks are $\alpha \sim 10^{-5} K^{-1}$ (Lauriello 1974, Richter and Simmons 1974).  Young's moduli $Y = (10 - 100) \times 10^9$ N m$^{-2}$ are typical for rock (Pariseau 2006).  If converted into kinetic energy with  efficiency $\eta$, thermal strain energy can generate fractured material with speeds (Jewitt and Li~2010)

\begin{equation}
v \sim \alpha \delta T \sqrt{\frac{\eta Y}{\rho}}.
\label{thermal}
\end{equation}

\noindent Substituting $\eta$ = 1 gives an upper bound to the ejection speeds of fracture fragments as $v \sim$ 20 m s$^{-1}$, for fracture at $\delta T$ = 1000 K.  This is sufficient to launch particles above the gravitational escape speed from asteroids up to radius $r \sim$ 20 km.  Thus, thermal fracture is a potential source of small particles and an agent capable of ejecting these particles from asteroids, but only those approaching the Sun very closely.  

\subsection{Thermal Dehydration}
Some carbonaceous chondrite classes (probable fragments of outer-belt asteroids)  contain 10\% - 20\% water by weight, bound into hydrated minerals (Jarosewich 1990).  If it could be liberated, this bound water might drive observable mass loss by entraining small grains in the gas flow produced as the water escapes into space.  

Laboratory experiments show that thermal dehydration of hydrous phyllosilicates of planetary relevance (e.g. serpentine, brucite, muscovite, talc) is characterized by activation energies in the range 325 - 400 kJ mol$^{-1}$ (Bose and Ganguly 1994).    Thermal dehydration is accompanied by a net volume change which can crack the dehydrating material and provide a source of small particles.  However, the above activation energies correspond to temperatures ($\sim$1000 K), much higher than normally found in the asteroid belt. For this reason, thermal dehydration is unlikely to play a role in a majority of the active asteroids since they have perihelia $\ge$2 AU and temperatures too low to trigger dehydration.

\subsection{Shock Dehydration}

Dehydration can be caused by shock pressure waves in impacts.   For example, shock dehydration of brucite (Mg(OH)$_2$) begins near 12 GPa and is complete by 56 GPa (50\% of the trapped water is lost at pressures 17 GPa) (Duffy et al.~1991).   Pressures this high are confined to a region  comparable in size to the projectile diameter, so that the volume of shock-dehydrated target material is small compared to the volume of the resulting impact crater.  Nevertheless, water released by impact could sustain mass loss from an asteroid for a timescale longer than that on which the impact ejecta dissipates.  For example, a 10 m radius projectile would dehydrate $\sim$10$^7$ kg of target.  With a water fraction of 10\%, some 10$^6$ kg of water could potentially leak from the impact site, enough to sustain a dusty flow at 0.1 kg s$^{-1}$ for $\sim$100 days.  
While shock dehydration must be considered, at best, a process of secondary importance, it does raise the prospect that gaseous products detected following an impact might, in the future, be misinterpreted as evidence for sublimating ice.

\subsection{Radiation Pressure Sweeping}
Dust particles can be swept from the surface of an asteroid through the action of radiation pressure.  To see this, we consider a grain located just above the surface and compare the gravitational acceleration towards the nucleus with the acceleration in the anti-sunward direction caused by radiation pressure,  the latter expressed as $\beta g_{\odot}$, where $g_{\odot}$ is the gravitational acceleration to the Sun.  A dust particle can be swept away by radiation pressure if

\begin{equation}
\beta \frac{G M_{\odot}}{R^2} > \frac{4}{3} \pi G \rho r
\label{amu0}
\end{equation}

\noindent in which we have assumed that the nucleus is at heliocentric distance $R$ and is spherical, of density $\rho$, radius $r$ and non-rotating.  Also, $G$ is the gravitational constant and $M_{\odot}$ is the mass of the Sun.  The dimensionless quantity $\beta$ depends upon the grain shape, composition and porosity but is principally a function of particle size, $a$.  As a useful first approximation, we take $\beta \sim 1/a_{\mu}$, where $a_{\mu}$ is the grain radius expressed in microns (Bohren and Huffman 1983).  Then, the condition for a grain to be lost to radiation pressure becomes

\begin{equation}
a_{\beta} < \frac{3 M_{\odot}}{4 \pi \rho r R^2}.
\label{amu1}
\end{equation}

\noindent  Substituting $\rho$ = 2000 kg m$^{-3}$ and expressing the nucleus radius in km and the heliocentric distance in AU, we find the critical radius for grain loss to be

\begin{equation}
a_{\beta} = 10 \left(\frac{1 km}{r}\right)\left(\frac{1 AU}{R}\right)^2.
\label{amu}
\end{equation}

\noindent  Equation (\ref{amu}) (also plotted in Figure \ref{critical}) shows that, for example, a 1 km radius object at $R$ = 3 AU would lose dust grains smaller than about $a \sim$ 1 $\mu$m.  Optically active particles ($a \ge$ 0.1 $\mu$m) can be swept away  throughout the asteroid belt provided their  parent bodies are smaller than about 10 km.  Larger particles can be lost if the nucleus is rotating and has an aspherical shape.

The above considerations are simplistic in that they take no account of the direction of the radiation pressure acceleration relative to the nucleus.  Grains released on the day-side, for instance, will be pushed back into the nucleus by radiation pressure from above.  On the other hand, grains released near the terminator will feel a net force in a direction determined by the vector sum of the local gravitational acceleration and radiation pressure acceleration. These grains have the potential to escape.  The principal limitation to efficacy of radiation pressure sweeping is, as with electrostatic launch, set by contact forces that hold small particles to the asteroid surface.  If these forces can be overcome, then radiation pressure sweeping can play a role across the asteroid belt for 10 km sized asteroids and smaller.

\section{Mechanisms for Individual Objects}

\subsection{133P and 238P}
The key observational property of these objects is that the activity is recurrent (Hsieh et al.~2004, Hsieh et al.~2010), qualitatively consistent with sublimation in the same way that mass loss from the classical comets is modulated by the varying insolation around the orbit.  Other mechanisms in Section (\ref{mechanisms}) are either inconsistent with recurrent mass loss (impact, rotational instability), or can be reconciled with it only in the most contrived way.  With equilibrium sublimation rates $\lesssim$(3 to 5)$\times$10$^{-5}$ kg m$^{-2}$ s$^{-1}$ at $R$ = 2.6 to 3.0 AU (Figure \ref{sublimation}), the most stringent production rates $\lesssim$0.01 kg s$^{-1}$ (Table \ref{physical})  imply mass loss from areas of exposed ice $\lesssim$200 to 300 m$^{2}$, corresponding to  circles of radius $\sim$10 m (Hsieh et al.~2004).  

The eccentricities of 133P and 238P are small (0.165 and 0.253, respectively) and, over the range of heliocentric distances swept, the expected sublimation mass flux varies by less than an order of magnitude (upper curve in Figure \ref{sublimation}).  This is a problem because a ten-fold reduction in activity at aphelion relative to perihelion would still produce detectable dust, whereas none is observed.  Instead, mass loss might be confined to a window near perihelion because of a shadowing effect of local topography on the nucleus, perhaps caused by recession of the sublimating surface, forming a pit (Equation \ref{recession}).   The surface recession rate at $R$ = 2.6 to 3.0 AU is (3 to 5)$\times$10$^{-8}$ m s$^{-1}$, so that a sublimation pit meters deep should develop in only a few orbits.  However, that the ice must still be close to the physical surface is shown by the appearance of coma near and soon after perihelion.  Ice more than a few centimeters deep would be thermally decoupled from the deposition of heat on the surface.

The evidence for ice sublimation, although self-consistent, remains indirect.   Detection of gaseous products of sublimation would provide  definitive evidence for the role of sublimation in active asteroids.  Just as in the classical comets, practical concerns render the daughter species (molecular fragments) optically more visible than the parent molecules sublimated from the ice.  The most prominent band in the optical spectra of comets is due to the trace species CN, itself a product of dissociation of the parent molecule HCN.  As noted in Section 2, several attempts to detect CN 3889\AA~in active asteroids have been reported, typically resulting in limits to the total gas production rate 0.1 - 1 kg s$^{-1}$,  about two to three orders of magnitude smaller than found in bright comets near 1 AU.

\begin{figure}
\epsscale{0.90}
\begin{center}
\plotone{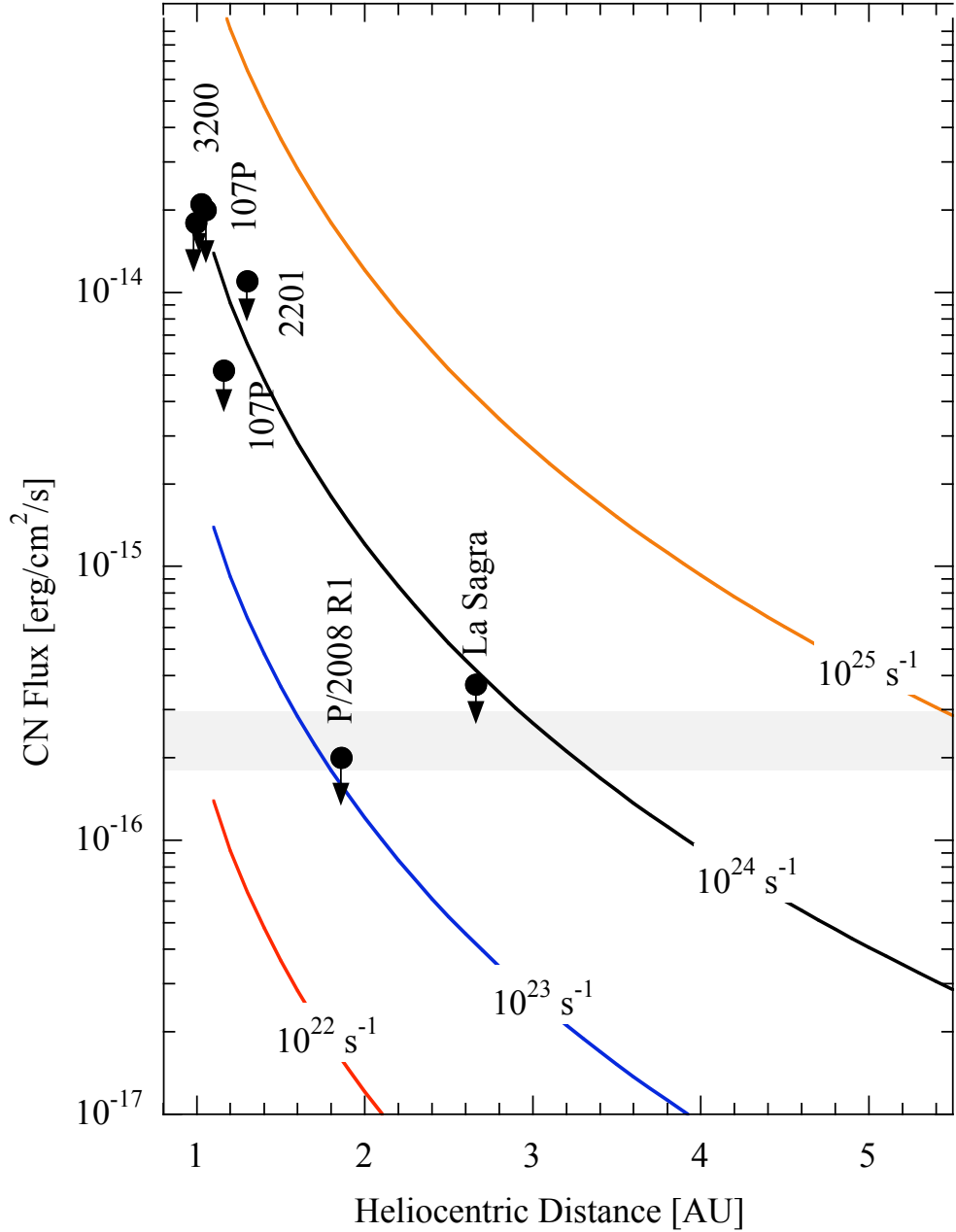}
\caption{Heliocentric distance dependence of the CN band flux for CN production rates $Q_{CN}$ = 10$^{22}$, 10$^{23}$, 10$^{24}$, and 10$^{25}$ s$^{-1}$.  The horizontal grey band shows the typical band sensitivity achievable using Keck in about an hour. Published limits to the CN flux are presented as black circles.  \label{cn_plot} } 
\end{center} 
\end{figure}

To set this on a quantitative basis, we show in Figure (\ref{cn_plot}) the expected CN band flux as a function of $R$ for several outgassing rates, as marked, computed as described in Jewitt et al.~(2009). The fluxes in the Figure were evaluated for opposition (i.e. $\Delta$ = $R$ - 1) and assume the Haser model and a spectroscopic slit 1$\arcsec \times$5$\arcsec$.   The Swings effect (in which the resonance fluorescence cross-section is affected by the heliocentric velocity through the Doppler effect) amounts to less than a factor of two, and has been neglected for simplicity.  Plotted on the figure are published limits to the CN band flux in active asteroids.  State of the art measurements, indicated by the horizontal grey band in the figure, reach CN fluxes $\sim$ (2 to 3)$\times$10$^{-16}$ erg cm$^{-2}$ s$^{-1}$, corresponding to $Q_{CN} \sim$ 10$^{23}$ s$^{-1}$ at $R$ = 2 AU but, by $R$ = 3 AU, the limits barely reach $Q_{CN} \sim$ 10$^{24}$ s$^{-1}$, showing how difficult is the spectroscopic detection of gas at larger distances.  The steep distance dependence is made worse by the rapidly declining sublimation rates from 2 AU to 3 AU (Figure \ref{sublimation}).   For the most favorable (high temperature) case in that figure (albedo 0.05) the sublimation rate drops by a factor of four from 2 AU to 3 AU.  For these reasons, spectroscopic detection of gas at main-belt distances is challenging even in many classical comets, where sublimation is the undoubted driver of the observed activity.  In so far as spectroscopic emission detections of gas in active asteroids are concerned, absence of evidence is not evidence of absence.

\subsection{(596) Scheila}
The brightening of Scheila was sudden, the scattering cross-section declined smoothly with time, and the morphology evolved in a way consistent with the expansion  of an impulsively ejected coma under the action of radiation pressure (Bodewits et al.~2011, Jewitt et al.~2011, Moreno et al.~2011, Hsieh et al.~2011, Ishiguro et al.~2011). All these signatures are consistent with impact production but are difficult or impossible to reconcile with the other mechanisms of Section (\ref{mechanisms}).  Scheila is also alone amongst the active asteroids, so far, in being very large (113 km diameter).  It presents a substantial cross-section for impact, but the resulting strong gravity precludes the action of electrostatic ejection, while the slow rotation eliminates any possibility of rotational instability.  Sublimation of water ice can eject small grains from Scheila, but the short coma fading time is difficult to understand given the sublimation rates in Figure (\ref{sublimation}).  

The outburst brightness was $\sim$1 mag brighter than the bare asteroid (Larson 2010, Jewitt et al.~2011).  Figure (\ref{dmag}) is consistent with the conclusion that the projectile was between 35 m and 60 m in diameter (Jewitt et al.~2011, Ishiguro et al.~2011 c.f. Bodewits et al.~2011
).  
\subsection{P/2010 A2}
\label{p2010a2}
Impact may account for the dust tail of P/2010 A2, but the interpretation in this case is non-unique. This body is comparatively tiny (only 120 m in diameter vs.~113 km for Scheila), with a tail mass suggesting collision with a projectile only a few meters in size (Jewitt et al.~2010, Snodgrass et al.~2010).  A key observational difference is that, whereas (596) Scheila faded on a timescale of one month, the tail of P/2010 A2 remained prominent $\sim$15 months after formation, despite the addition of no new dust (Jewitt et al.~2010, 2011).  This difference in fading timescales arises because the slowest particles to escape from a body leave the target with a speed comparable to the gravitational escape speed.  For  P/2010 A2 this is $\sim$0.1 m s$^{-1}$, allowing large, slow grains to persist for $>$1 yr after impact while, for Scheila, particles ejected at less than the  60 or 70 m s$^{-1}$ escape speed simply fell back on the surface (Jewitt et al.~2011, Ishiguro et al.~2011).  Furthermore, most grains traveling fast enough to be ejected were also small enough to be deflected by solar radiation pressure, providing another mechanism to quickly clear the coma. There are no direct measurements of the total, early-time brightening, only a limit $\Delta m <$ 19 (Jewitt et al.~2011).  However, the model suggests that if the P/2010 A2 were hit by a two meter-sized projectile, the brightening would be  $\Delta m \sim$ 15 (Figure \ref{dmag}), which is consistent with the empirical constraint.  Extreme brightening caused by impact has been suggested as an explanation of ``guest stars'' recorded in ancient Chinese records (Reach 1992). 

Rotational instability offers another explanation for mass loss from P/2010 A2 (Jewitt et al.~2010).  The small diameter of the primary body (120 m) corresponds to a short spin-up timescale under the action of YORP.   In fact, for sub-kilometer main-belt objects the YORP spin-up timescale is less than the timescale for collisional disruption  (Marzari et al.~2011, Jacobson and Scheeres 2011). While the rotation period of the primary body in P/2010 A2 is unknown, it would not be surprising to find that it had exceeded the centripetal limit.  Unfortunately, there are few published predictions for the detailed appearance of a rotationally disrupted body (Richardson et al.~2005).

\subsection{(3200) Phaethon}
Object (3200) Phaethon is distinguished from the other active asteroids by its small perihelion, $q$ = 0.14 AU,  allowing it to reach extreme sub-solar temperatures ($\sim$1000 K). These are sufficient to trigger thermal fracture and to dehydrate water-bearing minerals, if they are present (Jewitt and Li 2010).   It has been suggested that Phaethon was recently scattered inwards from a Pallas-like orbit ($a$ = 2.771 AU, $e$ = 0.281, $i$ = 33$\degr$; de Leon et al.~2010) where a large proportion of the asteroids are volatile-rich B- and C-types.  If Phaethon originated in the vicinity (or as part) of Pallas then finding hydrated minerals would not be surprising, since Pallas itself contains hydrated minerals.  The small asteroids associated dynamically with Phaethon (1999 YC and 2005 UD) should have similar compositions, but have not been observed close to perihelion and have shown no evidence for mass loss.

In addition to the creation of dust and fragments by thermal fracture and dehydration shrinkage, we conjecture that the escape of particles produced by these thermal mechanisms is facilitated by radiation pressure sweeping. This is especially important in (3200) Phaethon near perihelion relative to other asteroids.  According to Equation (\ref{amu}), particles up to 300 $\mu$m in diameter can be swept from its surface when at $q$ = 0.14 AU. Particles produced by thermal fracture or dehydration cracking, or lifted from the surface by electrostatic repulsion, would be rapidly swept away from it by radiation pressure (Jewitt and Li~2010).    

\subsection{Other Objects}
The mechanisms behind the mass loss in other active asteroids remain less clear.  Activity in 176P can be interpreted in terms of a sublimation model that is able to fit the observed coma morphology (Hsieh et al.~2010).  Models involving impulsive ejection (e.g. by impact) reportedly provide less good fits to the imaging data but, given the many unknown parameters in the models (e.g. the relation between dust particle size, scattering efficiency and speed)  the identification of sublimation in 176P cannot be regarded as definitive.  The same must be said for observations of P/2008 R1, where sublimation can fit the data (Jewitt et al.~2009) but where an impact origin may also apply.  We also regard the nature of activity in P/2008 R1 as unknown.  The coma isophotes of P/La Sagra are consistent with dust emission over an extended period (Moreno et al.~2011), as is integrated light photometry (after  correction for the unmeasured phase function of the comet, c.f. Hsieh et al.~2011).  Again, isophote models involve the assumption of numerous, unknown parameters and provide solutions which are non-unique. This is especially true for objects viewed at small angles to their orbital planes, as are most of the active asteroids.  Clear evidence of the weakness of models is the fact that Moreno et al.~(2010), using limited data, were able to fit the isophotes of P/2010 A2 with a model in which particles were released over an eight month period, inconsistent with the finding of an impulsive origin based on more and better imaging data (Jewitt et al.~2010, Snodgrass et al.~2010).    With this as a cautionary reminder, we regard the evidence concerning activity in 176P, P/2008 R1, P/La Sagra and 300163 as insufficient to diagnose the cause.  Evidence concerning 107P and (2201) Oljato is even more fragmentary, and the cause of their activity also unknown.

\begin{landscape}
\begin{figure}
\epsscale{0.95}
\begin{center}
\plotone{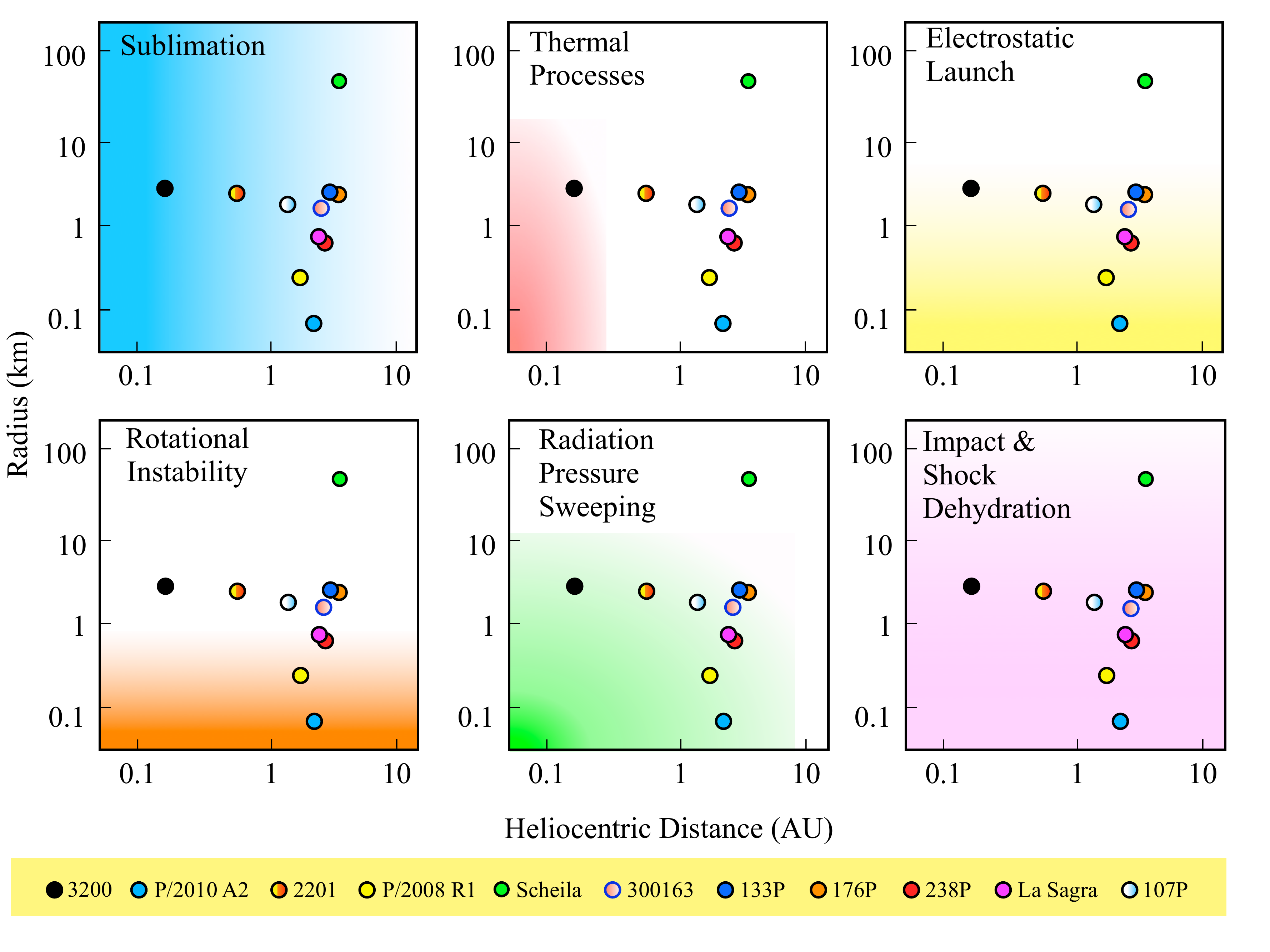}
\caption{Schematic diagrams of object radius vs.~heliocentric distance, with color fields marking the regions of action of the physical processes discussed in the text.  The color-coded circles denote the radii and distances at which mass loss has been reported in eleven active asteroids, identified in the key at the bottom.     \label{summary_figure} } 
\end{center} 
\end{figure}

\end{landscape}

The results of this and the previous section are summarized for convenience in Table (\ref{summary}) and shown schematically in Figure (\ref{summary_figure}).  The Figure shows the plane of  object radius, $r$, vs.~heliocentric distance, $R$, with the regions of action of the different physical processes considered in Section (\ref{mechanisms}) marked as color fields.  The eleven reported active asteroids are plotted at the heliocentric distances at which activity was observed.  The Figure shows that, for example,  (596) Scheila lies in a region of the $r$ vs.~$R$ plane in which only impact and sublimation are capable of ejecting particles.  (As described above, detailed studies of the morphology and time-dependence of the coma of Scheila show that impact is the true explanation).  Conversely, a majority of the active asteroids lie in regions of the $r$ vs.~$R$ plane where many processes are potentially important.  For example, in 238P, sublimation, electrostatic ejection, rotational instability, radiation pressure and impact process are all potentially active. Only through detailed physical investigation is it possible to discriminate between these possibilities (in favor of sublimation, in the case of 238P, based principally on the repetition of the observed mass-loss).   For many active asteroids, the physical observations needed to discriminate amongst mechanisms do not exist.

\section{Discussion}
\label{discussion}

Collisions are implicated in active asteroids both directly, as in the case of (596) Scheila and, perhaps, P/2010 A2, and indirectly as a trigger for activity (for example, to expose buried ice), as in 133P and 238P.  Here, we briefly examine the expected rate of collision between asteroids.

\begin{figure}
\epsscale{0.85}
\begin{center}
\plotone{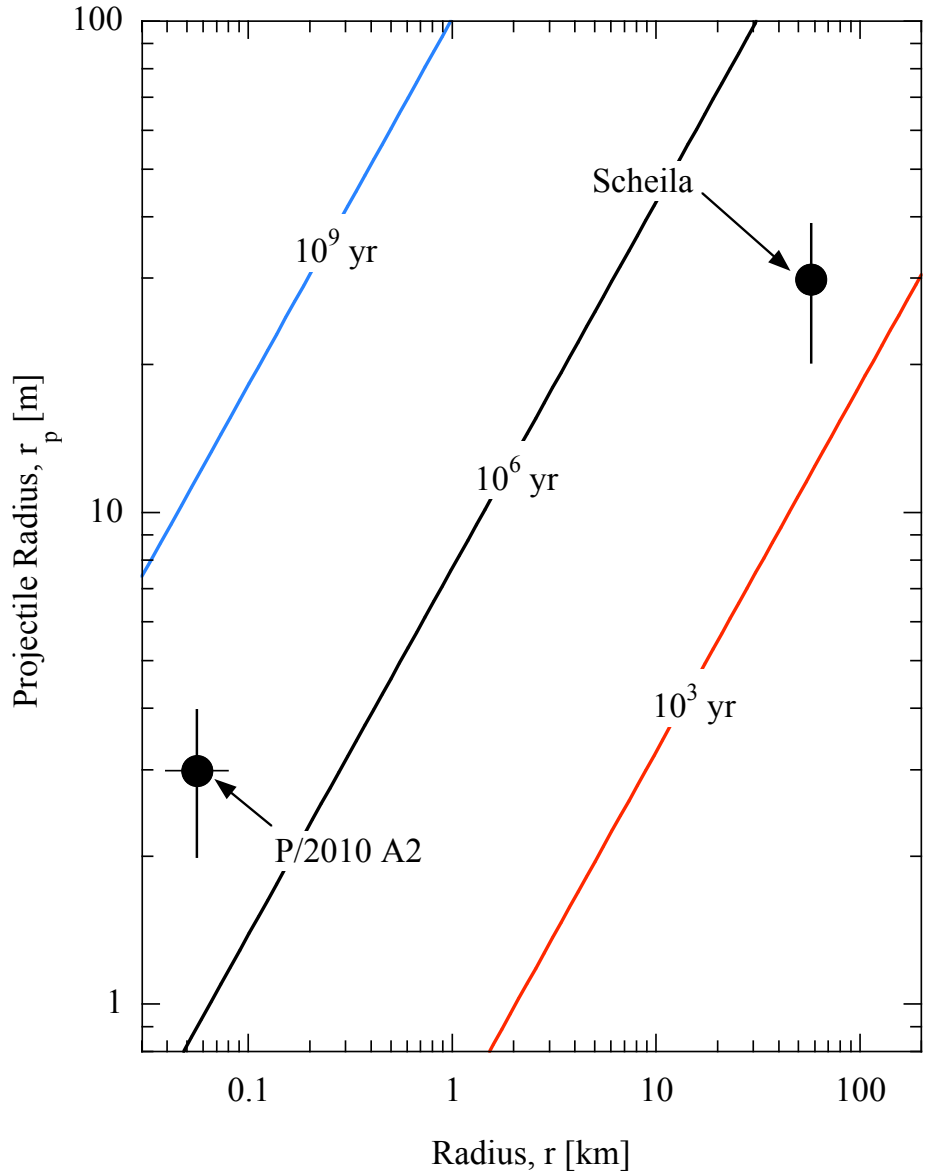}
\caption{Collision time, in years, as a function of target asteroid radius, for projectiles having radii 1 m to 100 m, as marked.  Lines of constant collision time  were computed from Equation (\ref{tauc2}).     Impact candidates P/2010 A2 and (596) Scheila  are shown with error bars reflecting independent estimates of the projectile radii, as discussed in the text.   \label{tauc_plot} } 
\end{center} 
\end{figure}

The typical collision probability per unit area in the asteroid belt is $P_c \sim$ 3$\times$10$^{-18}$ km$^{-2}$ yr$^{-1}$, with variations by a factor of several reflecting a collisional environment that varies with location in the belt  (Bottke et al.~1994).  The interval between impacts onto an asteroid of radius $r$ is 

\begin{equation}
\tau_c = \frac{1}{\pi r^2 P_c N_p (\ge r_p)}
\label{tauc}
\end{equation}

\noindent where $N( \ge r_p)$ is the number of impactors larger than $r_p$.  Estimates of the size distribution of the asteroids are many and varied, with significant uncertainties resulting from the unmeasured albedos of most asteroids, as well as from severe observational bias effects (Jedicke et al.~2002).  The uncertainties are particularly acute for sub-kilometer asteroids because such objects are faint and remain largely unobserved.  For radii $r >$ 1 km, the best-fitting differential power law index is about -2, albeit with significant, size-dependent deviations from this value.  For radii $r  \le$ 1 km, perhaps the best constraints on the distribution come  from the  impact crater size distribution on asteroid Gaspra.  There, craters from 0.4 km to 1.5 km in diameter (caused by projectiles perhaps 10 to 20 times smaller) are distributed as a power-law with a differential size index -3.7$\pm$0.5 (Belton et al.~1992; note that Chapman et al.~(1996) report that \textit{fresh} craters on Gaspra follow an even steeper distribution, with differential power law index -4.3$\pm$0.3).  We assume that the total main-belt  population is $\sim$1.4$\times$10$^{6}$ asteroids with diameters $>$1 km.  Combining these results and integrating over the size distribution we take the number of projectiles with radius $\ge r_p$ as

\begin{equation}
N_p (\ge r_p) = 2.6\times10^{13} \left(\frac{r_p}{1 m}\right)^{-2.7}
\label{N}
\end{equation}

\noindent where $r_p$ is expressed in meters.  Equations (\ref{tauc}) and (\ref{N}) together give

\begin{equation}
\tau_c = 4200 \left(\frac{r_p}{1 m}\right)^{2.7} \left(\frac{r}{1 km} \right)^{-2}  ~[yr]
\label{tauc2}
\end{equation}

\noindent  with $r$ in km.  Lines of constant collision time from Equation (\ref{tauc2}) are plotted in Figure (\ref{tauc_plot}). Impacted asteroid (596) Scheila is marked on Figure (\ref{tauc_plot}) with an error bar indicating different estimates of the projectile radius from Jewitt et al.~(2011; $a \sim$ 17 m) and Ishiguro et al.~(2011; $a \sim$ 40 m), respectively. The corresponding collision times are  3$\times$ 10$^{3}$ $\le \tau_c \le$3$\times$ 10$^{4}$ yr.  Given that there are $\sim$250 known asteroids as large as or larger than (596) Scheila, the mean interval between similar events is $\sim$(10 - 100) yr, statistically consistent with the detection of Scheila within the first decade of efficient, nearly real-time sky monitoring.  

Disrupted asteroid P/2010 A2 (interpreted as a $\sim$60 m radius object impacted by a projectile of characteristic size 2 to 4 meters; Jewitt et al.~2010) is also marked in Figure (\ref{tauc_plot}), indicating a collision time $\tau_c \sim$ 7 to 50$\times$10$^6$ yr.  With $N (\ge 60) \sim$ 4$\times$10$^8$ (Equation \ref{N}), the expected rate of similar events is $\sim$8 to 60 yr$^{-1}$.   Jewitt et al.~(2011) estimated the detection probability of P/2010 A2 clones as $\lesssim$6\%, so that of 8 to 60 similar events per year we would currently detect only $\sim$0.5 to 4. This is still higher than the actual rate of detection, counted as one object in perhaps a decade of efficient, nearly real-time sky monitoring by LINEAR and other survey telescopes.  The high rate could simply indicate the extreme uncertainty in using Equation (\ref{N}) to estimate the number of meter-sized projectiles. For example, the equation gives $N_p (\ge 5)$ = 3$\times$10$^{11}$, while published estimates of this number range from $\sim$10$^{9.5}$ to $\sim$10$^{12}$ (Davis et al.~2002), indicating a large uncertainty in $\tau_c$ at small projectile sizes.     Or it could indicate that activity in P/2010 A2 is caused by another process, possibly YORP spin-up, as remarked in Section (\ref{p2010a2}).  Further progress in understanding the impact rate in the asteroid belt hinges strongly upon better measurements of the sub-kilometer size distribution.     

Future wide-area sky surveys may help elucidate the mechanisms operating to cause asteroidal mass loss.  For example, we expect the spatial distribution of collisionally produced or triggered objects to be correlated with the regions of the asteroid belt in which the collision probability per unit time is highest.  Unfortunately, the published surveys for active asteroids so far are either biased (Hsieh 2009) or, if unbiased, detected no objects and so provide insufficient information (Gilbert and Wiegert 2009, 2010, Sonnett et al.~2011).  All but one of the known active asteroids were discovered serendipitously by a variety of surveys and methods, most too poorly quantified in terms of reported areal coverage and limiting magnitude to be used to interpret the spatial distribution.   This dismal situation deserves to be soon corrected by well-characterized, all-sky surveys having sensitivity sufficient to discover substantial numbers of active asteroids. 

\clearpage 

\section{Summary}
The number of mechanisms capable of producing mass loss from asteroids rivals the number of asteroids showing evidence for mass loss.  
No single mechanism can account for the varied examples of activity observed, but preferred explanations can be suggested for particular objects.

\begin{enumerate}

\item Sublimation of crystalline ice is effective to the outer edge of the asteroid belt and in asteroids up to a few 100 km in size.  Observational evidence for the sublimation of water ice is strongest in the two repeatedly active objects 133P/Elst-Pizarro and 238P/Read.  

\item Activity in (596) Scheila is unambiguously caused by the impact of a decameter-sized projectile, as indicated by the photometric and morphological evolution of this body.

\item Rotational instability is likewise broadly operable, but it is the sub-kilometer asteroids that are most likely to be driven towards instability by YORP effects.  The 120 m diameter P/2010 A2, if it is not also an impact relic, could be a rotationally disrupted body.

\item Thermal fracture and dehydration cracking are expected to supply dust particles only on asteroids that are very close to the sun ($\ll$1 AU) and smaller than about 20 km.  Radiation pressure sweeping can remove grains from 10 km asteroids up to  3 AU, but is more effective at smaller distances.  All these processes may operate in small perihelion object (3200) Phaethon.

\item Ejection of optically active grains (i.e.~$>$ 0.1 $\mu$m) by the electrostatic mechanism is possible for asteroids up to about 10 - 20 km in radius with an efficiency that is independent heliocentric distance.  No clear examples of this process have been identified.

\item The causes of mass loss from the other active asteroids cannot be reliably determined given the limited available data (see Table \ref{summary}).

\end{enumerate}

%
%
%
%
%
%
%
%
%
%

\acknowledgements
I thank Yan Fernandez, Henry Hsieh, Toshi Kasuga, Pedro Lacerda and Bin Yang for comments on the manuscript, and I appreciate support from NASA's Planetary Astronomy program.

\clearpage

\begin{deluxetable}{llllrcr}
\tablecaption{Summary of Orbital Properties
\label{orbital}}
\tablewidth{0pt}
\tablehead{
\colhead{Name} &\colhead{$T_J$\tablenotemark{a}} & \colhead{$a$\tablenotemark{b}} & \colhead{$e$  \tablenotemark{c}} & \colhead{$i$\tablenotemark{d}} & \colhead{$q$\tablenotemark{e}}   & \colhead{$Q$\tablenotemark{f}} }
\startdata

(3200) Phaethon & 4.508 & 1.271 & 0.890 & 22.17 & 0.140 & 2.402\\
P/2010 A2 & 3.582 & 2.291 & 0.124 & 5.26 & 2.007 & 2.575\\
(2201) Oljato & 3.299 & 2.172 & 0.713 & 2.52 & 0.623 & 3.721  \\
P/2008 R1 (Garradd) & 3.216 & 2.726 & 0.342 & 15.90 & 1.794 &3.658 \\
(596) Scheila & 3.208 & 2.928 & 0.165 & 14.66 & 2.445 & 3.411\\
300163 (2006 VW139) & 3.203 & 3.052 & 0.201 & 3.24 & 2.438 & 3.665 \\
133P/Elst-Pizarro & 3.184 &    3.157    & 0.165   &   1.39   &    2.636   & 3.678 \\
176P/LINEAR (118401) & 3.167 & 3.196 & 0.192 & 0.24 & 2.582 & 3.810 \\
238P/Read &  3.152 & 3.165 & 0.253 & 1.27 & 2.364 & 3.966 \\
P/2010 R2 (La Sagra) & 3.098 & 3.099 & 0.154 & 21.39 & 2.622 & 3.576\\
107P/Wilson-Harrington & 3.083 & 2.638 & 0.624 & 2.79  & 0.993 &  4.284\\

\enddata


\tablenotetext{a}{Tisserand parameter with respect to Jupiter}
\tablenotetext{b}{Semimajor axis [AU]}
\tablenotetext{c}{Orbital eccentricity}
\tablenotetext{d}{Orbital inclination}
\tablenotetext{e}{Perihelion distance [AU]}
\tablenotetext{f}{Aphelion distance [AU]}

\end{deluxetable}

\clearpage

\begin{landscape}
\begin{deluxetable}{llllllcrrr}
\tablecaption{Summary of Physical Properties
\label{physical}}
\tablewidth{0pt}
\tablehead{
\colhead{Name} &\colhead{$D$\tablenotemark{a}} & \colhead{$p_V$\tablenotemark{b}} & \colhead{$P$  \tablenotemark{c}} & \colhead{$B-V$  \tablenotemark{d}} & \colhead{$dm/dt$  \tablenotemark{e}}}
\startdata

(3200) Phaethon$^{1}$ & 5 - 7& 0.08 - 0.17 & 3.6 & 0.59$\pm$0.01 & N/A \\
P/2010 A2$^2$ & 0.12 & 0.1\tablenotemark{f} & ? & ? & N/A \\
(2201) Oljato$^{3}$ & 1.8 & 0.43$\pm$0.03 & ? & ? &  5? (gas)\\
P/2008 R1 (Garradd)$^4$ & $<$0.4 & 0.04\tablenotemark{f} & ? & 0.63$\pm$0.03 & $\le$1.5 (gas), 0.01 \\
(596) Scheila$^{5}$ & 113$\pm$2 & 0.038$\pm$0.004 & 15.848 & 0.71$\pm$0.03 & $\le$3 (gas) \\
300163 (2006 VW139)$^6$ & 3 & 0.04\tablenotemark{f} & ? & ? & ? \\
133P/Elst-Pizarro$^{7}$ & 3.8$\pm$0.6 &    0.05$\pm$0.02   & 3.471$\pm$0.001 & 0.65$\pm$0.03 & $<$0.04 (gas), 0.01, 0.7-1.6   \\
176P/LINEAR (118401)$^{8}$ & 4.0$\pm$0.4 & 0.06$\pm$0.02 & 22.23$\pm$0.01 & 0.63$\pm$0.02 & 0.1  \\
238P/Read$^9$ &  0.8 & 0.05\tablenotemark{f} & ? & 0.63$\pm$0.05 & 0.2\\
P/2010 R2 (La Sagra)$^{10}$ & 1.4 & 0.04\tablenotemark{f} & ? & ? & 4 \\
107P/Wilson-Harrington$^{11}$ &3.5$\pm$0.3 & 0.06$\pm$0.01 & 7.15& ? & $\le$150 (gas)  \\

\enddata


\tablenotetext{a}{Effective diameter (km)}
\tablenotetext{b}{Geometric albedo}
\tablenotetext{c}{Rotation period}
\tablenotetext{d}{Color index}
\tablenotetext{e}{Inferred mass loss rate in kg s$^{-1}$. Unless otherwise stated, the estimates are based on continuum measurements and refer to dust.  N/A means that no mass loss rate can be specified because the loss is not in steady state.}
\tablenotetext{f}{Value is assumed, not measured}
\tablenotetext{1}{Dundon 2005, $^2$Jewitt et al.~2010, $^{3}$ Tedesco et al.~(2004), Russell et al.~1984 $^4$Jewitt et al.~2009, $^5$Tedesco et al.~2002,  Warner 2006, $^6$Hsieh et al.~2011d, $^7$Hsieh et al.~2004, 2009a, 2011a, $^8$Hsieh et al.~2011, Licandro et al.~2011, $^9$ Hsieh et al.~2011c, $^{10}$Moreno et al.~2011b, Hsieh et al.~2011b,  $^{11}$Veeder et al.~1984, Fernandez et al.~1997,  Licandro et al.~2009, Urakawa et al. 2011.  }

\end{deluxetable}
\end{landscape}

\clearpage

\begin{deluxetable}{lccccc}
\tablecaption{Summary of Mechanisms\tablenotemark{a}
\label{summary}}
\tablewidth{0pt}
\tablehead{
\colhead{Name} &\colhead{Sublimation} & \colhead{Impact} & \colhead{Electrostatics} & \colhead{Rotation} & \colhead{Thermal} }
\startdata

(3200) Phaethon & $\times$ & ? & ? & ? & $\checkmark$   \\
P/2010 A2 &  $\times$  &  $\checkmark$   &  $\times$  &  $\checkmark$ &  $\times$  \\
(2201) Oljato &  ?  &   ?  &  ?  &   ?  &  $\times$  \\
P/2008 R1 (Garradd) &   ?   &   ?   &   ?   &   ?  &  $\times$  \\
(596) Scheila &  $\times$  &  $\checkmark$  &  $\times$  &  $\times$  &  $\times$   \\
300163 (2006 VW139) & ? & ? & ? & ? & $\times$ \\
133P/Elst-Pizarro & $\checkmark$ &    $\times$   & ? & ?   &   $\times$    \\
176P/LINEAR (118401) &  ?  &  ?  &  ?  &  $\times$  &  $\times$   \\
238P/Read &   $\checkmark$  &  $\times$  &  $\times$  &  ?  &  $\times$  \\
P/2010 R2 (La Sagra) &  ?  &  ?  &  ?  &  ?  &  $\times$   \\
107P/Wilson-Harrington & ? &  ?  &  ? &  $\times$  &  $\times$  \\

\enddata


\tablenotetext{a}{$\checkmark$ - evidence exists consistent with the process, $\times$ - evidence exists inconsistent with the process, $?$ - insufficient evidence exists }

\end{deluxetable}

\clearpage

\clearpage

\clearpage

\clearpage

\clearpage


\clearpage

\clearpage

\clearpage

\clearpage

\end{document}